\documentclass[printer]{aa}
\usepackage[utf8]{inputenc}

\usepackage{amsmath}
\usepackage{textcomp}
\usepackage{longtable}
\usepackage{xtab}
\usepackage{mathtools}
\usepackage{pdfpages}
\usepackage{subfig}
\usepackage{graphicx}
\usepackage{lscape}
\usepackage{txfonts}
\usepackage{natbib}
\bibpunct{(}{)}{;}{a}{}{,}
\usepackage{soul}
\usepackage{multirow}
\usepackage{xspace}
\usepackage{ulem}
\usepackage{tabularx}

\newcommand{\MSun}{\ensuremath{M_{\odot}}\xspace}
\newcommand{\RSun}{\ensuremath{R_{\odot}}\xspace}

\newcommand{\rhoSun}{\ensuremath{\rho_{\odot}}\xspace}
\newcommand{\MStar}{\ensuremath{M_{\star}}\xspace}
\newcommand{\RStar}{\ensuremath{R_{\star}}\xspace}
\newcommand{\rhoStar}{\ensuremath{\rho_{\star}}\xspace}
\newcommand{\MEarth}{\ensuremath{M_{\oplus}}\xspace}
\newcommand{\REarth}{\ensuremath{R_{\oplus}}\xspace}
\newcommand{\Rpl}{\ensuremath{R_{\mathrm{p}}}\xspace}
\newcommand{\Mpl}{\ensuremath{M_{\mathrm{p}}}\xspace}
\newcommand{\rhopl}{\ensuremath{\rho_{\mathrm{p}}}\xspace}
\newcommand{\rhoEarth}{\ensuremath{\rho_{\oplus}}\xspace}
\newcommand{\Teff}{\ensuremath{T_{\mathrm{eff}}}\xspace}
\newcommand{\Fbol}{\ensuremath{F_{\mathrm{bol}}}\xspace}
\newcommand{\logg}{\ensuremath{\log~(g)}\xspace}
\newcommand{\thetaUD}{\ensuremath{\theta_{\rm UD}}\xspace}
\newcommand{\thetaLD}{\ensuremath{\theta_{\rm LD}}\xspace}
\newcommand{\FeH}{[Fe/H]\xspace}

\newcommand{\TD}{\ensuremath{\Delta F}\xspace}

\newcommand{\dd}{{\rm d}}
\newcommand{\Ld}{\mathcal{L}}

\begin{document}

    \title{From the stellar properties of HD\,219134 to the internal compositions of its transiting exoplanets.}

  \author{R. Ligi\inst{1}, C. Dorn\inst{2}, A. Crida\inst{3,4}, Y. Lebreton\inst{5,6}, O. Creevey\inst{3}, F. Borsa\inst{1},  D. Mourard\inst{3}, N. Nardetto\inst{3}, I. Tallon-Bosc\inst{7}, F. Morand\inst{3}, E. Poretti\inst{1}.}

        \institute{INAF-Osservatorio Astronomico di Brera, Via E. Bianchi 46, I-23807 Merate, Italy \\
        \email{roxanne.ligi@inaf.it} 
        \and  
        University of Zurich, Institut of Computational Sciences, University of Zurich, Winterthurerstrasse 190, CH-8057, Zurich, Switzerland
        \and
        Universit\'e C\^ote d'Azur, Observatoire de la C\^ote d'Azur, CNRS, Laboratoire Lagrange, Bd de l'Observatoire, CS 34229, 06304 Nice cedex 4, France
         \and
         Institut Universitaire de France, 103 Boulevard Saint-Michel, 75005 Paris, France
         \and
         LESIA,  Observatoire  de  Paris,  PSL  Research  University,  CNRS, Sorbonne Universit\'es, UPMC Univ. Paris 06, Univ. Paris Diderot, Sorbonne Paris Cit\'e, 92195 Meudon Cedex, France
         \and
           Univ. Rennes, CNRS, IPR (Institut de Physique de Rennes) - UMR 6251, F-35000 Rennes, France
                 \and
                Univ. Lyon, Univ. Lyon1, Ens de Lyon, CNRS, Centre de Recherche Astrophysique de Lyon UMR5574, F-69230 Saint-Genis-Laval, France
        }       

   \date{Received 5 July 2019 / Accepted 20 September 2019}


  \abstract
    {The harvest of exoplanet discoveries has opened the area of exoplanet characterisation. But this cannot be achieved without a careful analysis of the host star parameters.}
   {The system of HD\,219134 hosts two transiting exoplanets and at least two additional non-transiting exoplanets. We revisit the properties of this system using direct measurements of the stellar parameters to investigate the composition of the two transiting exoplanets.}
   {We used the VEGA/CHARA interferometer to measure the angular diameter of HD\,219134. We also derived the stellar density from the transits light curves, which finally gives a direct estimate of the mass.
   This allowed us to infer the mass, radius, and density of the two transiting exoplanets of the system. We then used an inference model to obtain the internal parameters of these two transiting exoplanets. }
   {We measure a stellar radius, density, and mass of \RStar$=0.726\pm0.014$ \RSun, \rhoStar$=1.82\pm0.19$ \rhoSun, and \MStar$=0.696\pm0.078$ \MSun, respectively; there is a correlation of 0.46 between \RStar and \MStar. This new mass is lower than that derived from the C2kSMO stellar evolutionary model, which provides a mass range of 0.755$-$0.810 ($\pm 0.040$) \MSun.
   Moreover, we find that planet $b$ and $c$ have smaller radii than previously estimated of $1.500\pm0.057$ and $1.415\pm0.049$ \REarth respectively; this clearly puts these planets out of the gap in the exoplanetary radii distribution and validates their super-Earth nature. Planet $b$ is more massive than planet $c$, but the former is possibly less dense. We investigate whether this could be caused by partial melting of the mantle and find that tidal heating due to non-zero eccentricity of planet $b$ may be powerful enough.
   }
        {The system of HD\,219134 constitutes a very valuable benchmark for both stellar physics and exoplanetary science. The characterisation of the stellar hosts, and in particular the direct determination of the stellar density, radius, and mass, should be more extensively applied to provide accurate exoplanets properties and calibrate stellar models. }

               
\authorrunning{R. Ligi}
\titlerunning{Stellar and planetary properties of HD\,219134.}
   \maketitle
%

\section{Introduction}
\label{sec:Introduction}

The huge harvest of exoplanets discovered by the space telescopes \textit{Kepler} \citep{Borucky2010} and CoRoT \citep{Baglin2003} has led to the understanding that exoplanets are the rule rather than the exception. We have now moved to the era of exoplanet characterisation, and the next challenge is to understand how common rocky planets are and if any are suitable for life. 
The most interesting exoplanets to study are certainly the transiting exoplanets, as the transit light curve allows us to know the planetary radius. An additional radial velocity (RV) follow-up provides the planetary mass and thus the planetary density. The three ingredients to estimate planetary bulk composition are then gathered. But this is only true if the stellar radius and mass are known. Up to now, most of transiting exoplanet hosts have been very faint, driven by the search for exoplanets rather than their characterisation, often leading to inaccurate and/or imprecise stellar parameters. This makes the characterisation of the whole exoplanetary system difficult and the determination of the exoplanetary internal structure approximate.

Several methods can be employed to obtain the stellar parameters. Concerning the mass, it is often determined indirectly, as only stars in binary systems can have their mass directly measured if the system inclination is known. However, if an exoplanet is transiting its host star, the density of the star can be directly inferred from the transit light curve \citep{Seager2003}. Then, in the case of bright stars, the radius can be directly determined using interferometry, which is a high angular resolution technique aimed at measuring the angular diameter of stars with a precision up to a few percent \citep[][e.g.]{Baines2010,Boyajian2012a,Boyajian2012b,Huber2012,Creevey2012, Ligi2012, Creevey2015, Ligi2016}. The mass can thus be directly computed from the transit and interferometric measurements.
This method has recently been used by \cite{Crida2018, Crida2018RNAAS} to derive the mass of the very bright star 55~Cnc with a precision of 6.6$\%$ using the interferometric diameter measured by \cite{Ligi2016} and the density from the transit light curve obtained for 55~Cnc $e$ \citep{Bourrier2018}. This yielded the best characterisation of the transiting super-Earth 55~Cnc $e$ so far and a new estimate of its internal composition.

HD\,219134 (HIP\,114622, GJ\,892) is also a bright (V=5.57) K3V star 6.5 parsecs away from us. \cite{Motalebi2015} first detected four exoplanets around the star from RV measurements using the High Accuracy Radial velocity Planet Searcher for the Northern hemisphere (HARPS-N) on the Telescopio Nazionale Galileo (TNG). Moreover, \textit{Spitzer} time-series photometric observations allowed the detection of the transit of planet $b$, leading to the estimate of a rocky composition. The same year, \cite{Vogt2015} claimed the detection of six planets around HD\,219134 from the analysis of RV obtained with the HIgh Resolution Echelle Spectrometer (HIRES) on Keck I Observatory and the Levy Spectrograph at the Automated Planet Finder Telescope (Lick Observatory). These authors derived similar periods for planets $b$, $c,$ and $d$, the other diverging because of the different Keplerian analysis of the RV signal leading to a different number of planets.
Later, \cite{Gillon2017} reported additional \textit{Spitzer} observations of the system that led to the discovery of the transit of the second innermost planet, HD\,219134~$c$. The two innermost planets seem rocky, but more interestingly, planet $c$ shows a higher density while it has a lower mass than planet $b$. The detailed planetary data and their relative differences place additional constraints on their interiors with implications to their formation and evolution.

In this paper, we report new observations of HD\,219134 using the Visible spEctroGraph and polArimeter (VEGA) instrument on the Center for High Angular Resolution Astronomy (CHARA) interferometric array that led to a new accurate determination of angular diameter of this star (Sect.~\ref{sec:Interfero}).
In Sect.~\ref{sec:StellarParam}, we determine the stellar radius and density, and derive the joint probability density function (PDF) of the stellar mass and radius independently of stellar models.
We then use the PDF to compute the new parameters of the two transiting exoplanets and we revisit those of the non-transiting exoplanets in Sect.~\ref{sec:RDMplanets}. Finally, we derive the internal composition of planets $b$ and $c$ in Sect. \ref{sec:InternalComposition} using a planetary interior model, and we discuss the possible cause of the different densities of planets $b$ and $c$ in Sect. \ref{sec:simus}. We conclude in Sect. \ref{sec:Conclusion}.

\section{Interferometric measurement of the angular diameter with VEGA/CHARA}
\label{sec:Interfero}

We used the technique of interferometry to measure the angular diameter of HD\,219134. These measurements constitute the first step to determine the other fundamental parameters of this star.

\subsection{Observations and data reduction}

We observed HD\,219134 from 2016 to 2018 using the VEGA/CHARA instrument at visible wavelengths (see Table~\ref{tab:Obs}) and medium resolution. The spectro-interferometer VEGA \citep{Mourard2009, Ligi2013} is based on the CHARA array \citep{tenBrummelaar2005}, which takes advantage of the six 1 m telescopes distributed in a Y-shape to insure wide (\textit{u,v}) coverage. It can be used at medium (5000) or high spectral resolution (30 000) and with baselines ranging from 34 to 331 m in the two telescope (2T), 3T, or 4T modes.
The observations were calibrated following the sequence calibrator - science star - calibrator, and were performed using different configurations (Table \ref{tab:Obs}), mainly in the 2T mode at once to optimise the signal-to-noise ratio (S/N) of the observations. The calibrator stars were selected into the SearchCal software\footnote{\texttt{http://www.jmmc.fr/searchcal$\_$page.htm}} (Table~\ref{tab:calibs}), and we used the uniform disc diameter in the $R$ band  (UDDR) found in the JSDC2 \citep{Bourges2014} or SearchCal \citep{Chelli2016} catalogue otherwise. However, for conservative reasons, we decided to use an uncertainty of $7\%$ or that given in the JSDC1 \citep{Bonneau2006} if higher. We selected the calibrators with several criteria: in the neighbourhood of the star, discarding variable stars and multiple systems, and with high squared visibilities, allowing an optimal measurement of the instrumental transfer function.
Finally, the data were reduced using the \texttt{vegadrs} pipeline \citep{Mourard2009, Mourard2011} developed at Observatoire de la C\^ote d'Azur.  For each observation, we selected two non-redundant spectral bands of 20~nm wide centred at 685 nm, 705 nm, or 725 nm in most cases to derive the squared visibility ($V^2$), but the reddest band is sometimes of bad quality or features absorption lines and cannot be used. In total, we collected 36 data points, which are shown in Fig.~\ref{fig:visibilities}.

\begin{table*}
\centering
\caption{Observing log.}
\begin{tabular}{l l l c c c c c c c c}
\hline \hline
\multicolumn{2}{c}{Date}        &       Telescopes      &       Bas. length     &       Seq.    &       S/N     &       V$^2$   &       $\sigma V^2_{\rm Stat}$ &       $\sigma V^2_{\rm Syst}  $ &     $\lambda$               & $\delta\lambda$ \\
MJD     &        UT     &               &       [m]     &               &               &               &               &                &               [nm]    & [nm]  \\
\hline
57621.451       &       2016-08-21      &       W2E2    &       105.30  &       C1 – sci – C1  &       4.1     &       0.206   &       0.050   &       1.43E-03        &       685     &       20      \\
57711.177       &       2016-11-19      &       E1E2    &       65.79   &       C1 – sci – C1  &       13.9    &       0.694   &       0.050   &       1.78E-03        &       705     &       20      \\
57711.177       &               &               &       65.79   &               &       15.0    &       0.751   &       0.050   &       1.76E-03        &       725     &       20      \\
57711.269       &       2016-11-19      &       E1E2    &       65.48   &       C1 – sci – C1  &       10.9    &       0.546   &       0.050   &       1.41E-03        &       705     &       20      \\
57711.292       &       2016-11-19      &       E1E2&   65.36   &       C1 – sci – C1  &       11.3    &       0.567   &       0.050   &       1.69E-03        &       705     &       20      \\
57712.150       &       2016-11-20      &       W1W2    &       107.82  &       C1 – sci – C2  &       5.6     &       0.280   &       0.050   &       1.48E-03        &       705     &       20      \\
57712.172       &       2016-11-20      &       W1W2    &       107.20  &        C2 – sci – C1       &       6.0     &       0.300   &       0.050   &       1.64E-03        &       705     &       20      \\
57712.172       &               &               &       107.20  &                &       2.2     &       0.111   &       0.050   &       9.24E-04        &       725     &       20      \\
57712.221       &       2016-11-20      &       W1W2    &       103.69  &       C1 – sci – C2  &       3.8     &       0.189   &       0.050   &       9.39E-04        &       705     &       20      \\
57712.221       &               &               &       103.69  &               &       6.1     &       0.304   &       0.050   &       1.46E-03        &       725     &       20      \\
57712.245       &       2016-11-20      &       W1W2    &       101.20  &        C2 – sci – C1       &       4.9     &       0.245   &       0.050   &       1.16E-03        &       705     &       20      \\
57712.245       &               &               &       101.20  &               &       5.3     &       0.267   &       0.050   &       1.21E-03        &       725     &       20      \\
57734.160       &       2016-12-12      &       S1S2    &       29.52   &       C1 – sci – C1  &       11.6    &       0.987   &       0.085   &       5.80E-04        &       705     &       20      \\
57734.160       &               &               &       29.52   &               &       11.6    &       0.958   &       0.083   &       5.38E-04        &       725     &       20      \\
57960.258       &       2017-07-26      &       W1W2    &       95.26   &       C3 – sci – C3  &       5.9     &       0.296   &       0.050   &       1.63E-03        &       705     &       20      \\
57960.258       &               &               &       95.26   &               &       5.8     &       0.289   &       0.050   &       1.51E-03        &       725     &       20      \\
57960.279       &       2017-07-26      &       W1W2    &       96.32   &       C3 - sci – C4    &       6.1     &       0.304   &       0.050   &       2.14E-03        &       705     &       20      \\
57960.279       &               &               &       96.32   &               &       6.7     &       0.334   &       0.050   &       2.28E-03        &       725     &       20      \\
57960.302       &       2017-07-26      &       W1W2    &       97.91   &       C4 - sci – C4    &       6.6     &       0.331   &       0.050   &       2.57E-03        &       705     &       20      \\
57960.302       &               &               &       97.91   &               &       7.5     &       0.373   &       0.050   &       2.74E-03        &       725     &       20      \\
57960.324       &       2017-07-26      &       W1W2    &       99.79   &       C4 - sci – C3    &       5.8     &       0.290   &       0.050   &       2.07E-03        &       705     &       20      \\
57960.324       &               &               &       99.79   &               &       5.9     &       0.294   &       0.050   &       2.02E-03        &       725     &       20      \\
57964.369       &       2017-07-30      &       E1E2    &       61.12   &       C4 - sci – C3    &       9.8     &       0.489   &       0.050   &       1.06E-03        &       705     &       20      \\
57964.369       &               &               &       61.12   &               &       9.1     &       0.455   &       0.050   &       9.33E-04        &       725     &       20      \\
57964.510       &       2017-07-30      &       E1E2    &       65.87   &       C3 – sci – C3  &       15.0    &       0.750   &       0.050   &       2.25E-03        &       705     &       20      \\
57964.510       &               &               &       65.87   &               &       16.5    &       0.823   &       0.050   &       2.18E-03        &       725     &       20      \\
57965.343       &       2017-07-31      &       E1E2    &       59.00   &       C3 – sci – C3  &       14.7    &       0.737   &       0.050   &       2.19E-03        &       705     &       20      \\
57965.343       &               &               &       59.00   &               &       14.9    &       0.745   &       0.050   &       1.79E-03        &       725     &       20      \\
58299.481       &       2018-06-30      &       E1E2    &       63.21   &       C3 – sci - C3     &      10.5    &       0.525   &       0.050   &       1.24E-03        &       703     &       20      \\
58299.481       &               &               &       63.21   &               &       11.5    &       0.577   &       0.050   &       1.29E-03        &       723     &       20      \\
58299.501       &       2018-06-30      &       E1E2    &       64.25   &       C3 – sci - C3    &       13.2    &       0.659   &       0.050   &       1.58E-03        &       703     &       20      \\
58299.501       &               &               &       64.25   &               &       13.3    &       0.665   &       0.050   &       1.50E-03        &       723     &       20      \\
58302.334       &       2018-07-03      &       W1W2    &       95.87   &       C3 – sci – C4  &       7.6     &       0.378   &       0.050   &       2.77E-03        &       703     &       20      \\
58302.334       &               &               &       95.87   &               &       6.3     &       0.361   &       0.057   &       2.21E-03        &       723     &       20      \\
58302.351       &       2018-07-03      &       W1W2    &       96.91   &       C4 - sci - C3      &       3.7     &       0.220   &       0.060   &       1.47E-03        &       703     &       20      \\
58302.351       &               &               &       96.91   &               &       7.7     &       0.385   &       0.050   &       2.41E-03        &       723     &       20      \\
\hline
\end{tabular}
\tablefoot{From left to right, the table shows the observing date, telescopes used, projected baseline lengths, observing sequence (``sci'' refers to the science target, and ``C-'' to the calibrator; see Table \ref{tab:calibs}), the S/N, measured squared visibility, statistical and systematic errors, observation wavelength, and corresponding bandwidth.}
\label{tab:Obs}
\end{table*}

\begin{table}
\caption{Angular diameters of the calibrators used.}
\begin{tabular}{l l c c}
\hline \hline
Cal. & Name &  $\theta_{\rm UD} \pm \sigma \theta_{\rm UD}$ [mas] & Ref. \\
\hline
C1 & HD\,1279   &  $0.183\pm0.013$  & (1) \\
C2 & HD\,209419 & $0.158\pm0.011$ & (1) \\
C3 & HD\,218376 & $0.188\pm0.013$ & (2)\\
C4 & HD\,205139 & $0.174\pm0.017$ & (2) \\
\hline
\end{tabular}
\tablebib{(1) JSDC2 \citep{Bourges2014}\,; (2) SearchCal \citep{Chelli2016}.}
\label{tab:calibs}
\end{table}

\subsection{Angular diameter}
\label{subsec:angdiam}

\begin{figure}
\includegraphics[scale=0.5]{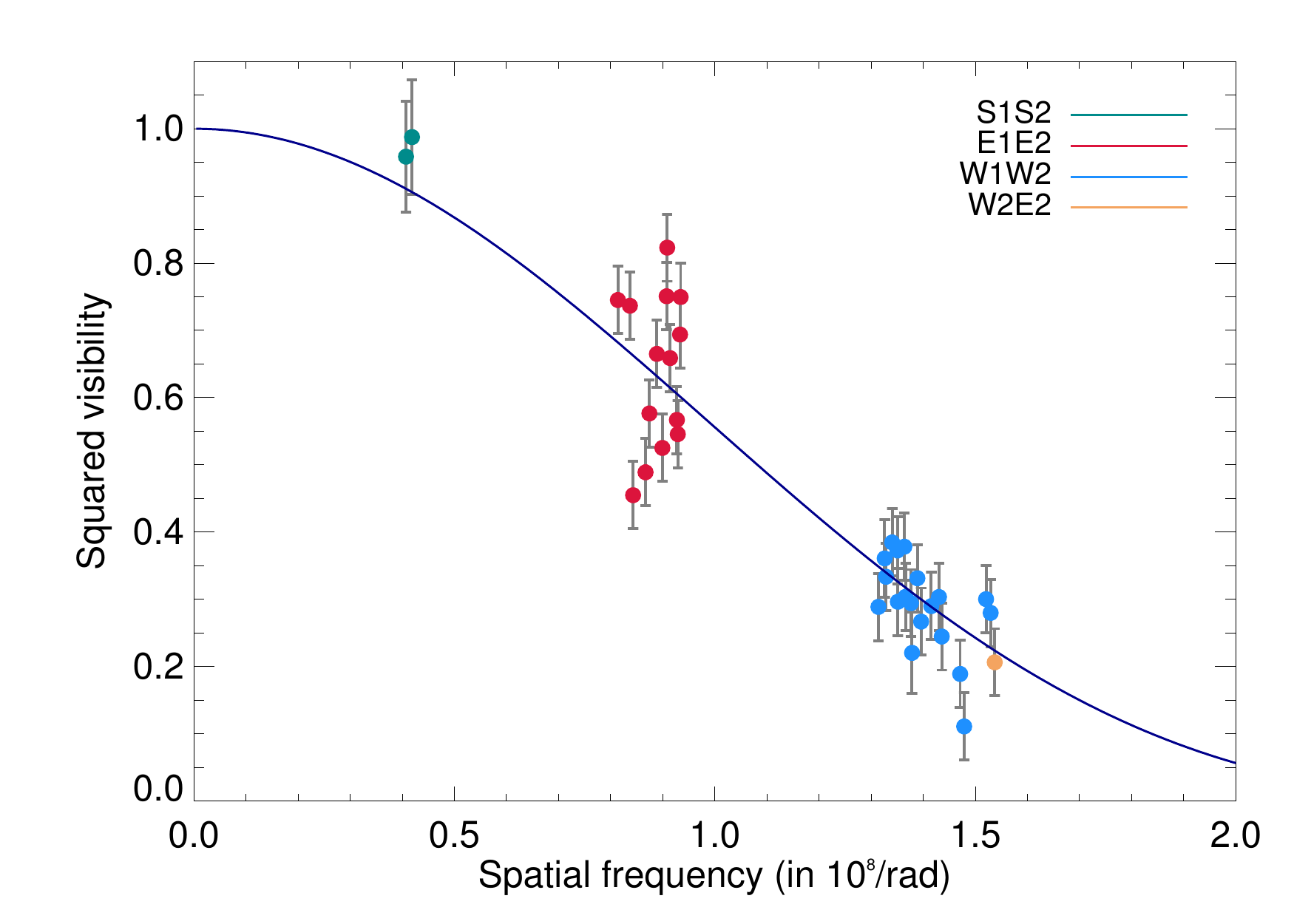}
\caption{Squared visibilities obtained with VEGA/CHARA for HD\,219134. The different colours represent the data points obtained with different baselines. The solid line represents the model of LD diameter.}
\label{fig:visibilities}
\end{figure}

The squared visibilities that we obtained (Fig.~\ref{fig:visibilities}, coloured filled circles) are well spread on the $V^2$ curve. We note some dispersion around $0.7\times10^8/$rad (corresponding to the E1E2 configuration) but it is taken into account in the computation of the error on the angular diameter. We also adopted a conservative approach by setting a minimum error of $5\%$ on $V^2$ to balance the known possible bias with VEGA \citep{Mourard2012,Mourard2015}.

We used the \texttt{LITpro} software \citep{Tallon2008} to fit our visibility points and derive the angular diameter of HD\,219134 and its related uncertainty. Taking a model of uniform disc, we obtained \thetaUD $=0.980\pm0.020$~millisecond of arc (mas\,; Table~\ref{tab:StellarParam}). 
However, this simple representation is not realistic and we thus used a linear limb-darkening (LD) model to refine it, as the LD diameter (\thetaLD) cannot be directly measured. We must indeed use empirical tables of LD coefficients $\mu_\lambda$, which depend on the effective temperature \Teff, gravity \logg, and metallicity \FeH at a given wavelength $\lambda$. We used \cite{Claret2011} tables as a start in the $R$ and $I$ band since we observed between 685 and 720 nm, and proceeded on interpolations to obtain a reliable LD coefficient at our wavelength, as described in \cite{Ligi2016}.
The LD coefficients in \cite{Claret2011} tables are given in steps of $250$~K for \Teff, 0.5 dex for \logg and less uniform steps for \FeH. We set a starting value of these parameters to perform our interpolation in between the surrounding values. We searched in the literature previous values of \logg and \FeH through the SIMBAD database\footnote{Available at \texttt{http://cdsweb.u-strasbg.fr/}} and calculated the median and standard deviation of the values given there (see selected values in Table~\ref{tab:ParamLit} and the medians in Table \ref{tab:StellarParam}). Beforehand, we eliminated aberrant values and values obtained before the year 2000, to insure recent and probably more reliable estimates. Since many values of the metallicity could be derived from a same data set, and because the uncertainty in the various papers can be higher than our standard deviation (0.05 dex), we set the uncertainty on \FeH to 0.1 dex. Concerning the starting \Teff value, we used that fitted through the spectral energy distribution (SED; $T_{\rm eff, SED} = 4\,839$ K, Sect.~\ref{subsec:SED}). Since the star is close by (distance, $d = 6.533\pm0.038$ pc, Table \ref{tab:StellarParam}), we set the reddening to $A_{\rm v} = 0.0\pm0.01$ mag. This value is consistent with the extinction given by the Stilism \citep{Lallement2014} 3D map of the galactic interstellar matter (E(B-V) $= 0\pm0.014$) but corresponds to a smaller uncertainty on the extinction (0.0034 mag).

For each filter, we first computed the linear interpolation of the LD coefficients corresponding to the surrounding values of \FeH, \logg and \Teff of our star. We then averaged the two coefficients coming out from each filter to get a final coefficient. 
Then, we used the \texttt{LITpro} software to fit our data using a linear LD model while fixing in the model our new LD coefficient. 
This results in \thetaLD $=1.035 \pm 0.021$~mas (2$\%$ precision). It has to be noted that using different LD laws does not significantly change the final diameter as we are not sensitive to it in the first lobe of visibility. 
If we set \Teff$= 4\,750$ K, \logg= 4.5 dex, and \FeH = 0.1 dex, a quadratic LD law described by \cite{Claret2011} yields \thetaLD$=1.047\pm0.022$~mas in the $R$ band (using the LD coefficients $a_{1,R}=0.5850$ and $b_{2,R}=0.1393$ given in the table) and \thetaLD$=1.033\pm0.022$~mas in the $I$ band (taking $a_{1,I}=0.4490$ and $b_{2,I}=0.1828$). Similarly, averaging $a_{1,R}$ and $a_{1,I}$ coefficients on the one hand, and $b_{2,R}$ and $b_{2,I}$ on the other hand, leads to \thetaLD $=1.040\pm0.022$~mas, and thus a value within the error bars of our first estimate.
Our determined angular diameter is smaller than that previously measured with the CHARA Classic beam combiner \citep[$1.106\pm0.007$~mas;][]{Boyajian2012b}. 
Although their visibilities seem more precise, we stress that we obtain higher spatial frequency data, which resolves the star better.
The angular diameter derived from the SED $\theta_{\rm SED}$ is also very consistent with our measurement (1.04 mas, see Sect. \ref{subsec:SED}).  

\section{Stellar parameters}
\label{sec:StellarParam}

The new angular diameter constitutes the basis of our analysis. It is now possible to determine the other stellar parameters from our interferometric measurements, and to compare these parameters with those derived from stellar evolution models.

\begin{table}
\caption{Stellar parameters of HD\,219134.}
\begin{tabularx}{\textwidth /2}{lcl}
\hline \hline
Parameter & Value & Ref \\
\hline
\multicolumn{3}{c}{\textit{Coordinates and photometry}} \\
\hline
RA (J2000) &    23$^h$13$^m$16$^s$.97   &  \textit{Gaia} DR2$^{(1)}$\\
DEC (J2000)  &  +57$^{\degr}$10$^{\arcmin}$06$^{\arcsec}$.08    &  \textit{Gaia} DR2$^{(1)}$ \\ 
$\pi$ [mas] &   $153.081 \pm 0.0895$    &  \textit{Gaia} DR2$^{(1)}$ \\
$d$ [pc] &      $6.533 \pm 0.038$ & \textit{Gaia} DR2$^{(1),(a)}$   \\ 
V [mag] &        $5.570 \pm 0.009$      & CDS$^{(2)}$ \\
K [mag]  &      $3.25 \pm 0.01$ & CDS$^{(2)}$ \\
$L_\star$ [$L_{\sun}$] &        $0.30$  &  \textit{Gaia} DR2$^{(1)}$  \\
\hline
\multicolumn{3}{c}{\textit{Interferometric parameters}} \\
\hline
\thetaUD        [mas] & $0.980 \pm 0.020$ (2$\%$) &  This work, Sect. \ref{subsec:angdiam} \\
$u_{\lambda}$           &       $0.588$         &   This work$^{(b)}$, Sect. \ref{subsec:angdiam}   \\ 
\thetaLD        [mas] & $1.035 \pm0 .021$ (2$\%$) &  This work, Sect. \ref{subsec:angdiam} \\ 
\hline
\multicolumn{3}{c}{\textit{Fixed parameters}} \\
\hline
A$_{\rm v}$ [mag]  & $0.0 \pm 0.01$     & This work, Sect. \ref{subsec:SED} \\
$\rm [Fe/H]$ & $0.07 \pm 0.1$ & SIMBAD$^{(c)}$, Sect. \ref{subsec:angdiam}  \\  
$\log(g$ [cm\,s$^{-2}$]$)$  & $4.57 \pm 0.14$ & SIMBAD$^{(c)}$, Sect. \ref{subsec:angdiam} \\ 
 \hline
 \multicolumn{3}{c}{\textit{Fitted parameters}} \\
 \hline
$T_{\rm eff, SED}$ [K] & $4\,839 \pm 25$  &  This work, Sect. \ref{subsec:SED} \\
$\theta_{\rm SED}$ [mas] &  $1.043 \pm 0.013$   &  This work, Sect. \ref{subsec:SED} \\ 
$F_\mathrm{bol}$ [erg\,s$^{-1}$\,cm$^{-2}$] & \ \hspace{-8pt} $(19.86 \pm 0.21)\cdot 10^{-8}$ \hspace{-8pt}   \   &  This work, Sect. \ref{subsec:SED} \\
 \hline
 \multicolumn{3}{c}{\textit{Measured and computed parameters}} \\
 \hline
$\Teff$ [K] & $4\,858 \pm 50$ & This work, Sect. \ref{subsec:SED} \\
$R_\star$ [$R_{\sun}$]   &      $0.726 \pm 0.014$       &  This work, Sect. \ref{sec:RDM} \\ 
$L_\star$ [$L_{\sun}$] &        $0.264 \pm 0.004$       &  This work, Sect. \ref{subsec:SED} \\
$\rho_\star$ [$\rho_{\sun}$] &$1.82 \pm 0.19$   &  This work, Sect. \ref{sec:RDM} \\
$M_\mathrm{\star}$ [$M_{\sun}$]  &     $0.696 \pm 0.078$     & This work$^{(d)}$, Sect. \ref{sec:RDM}   \\ 
$M_\mathrm{grav,\star}$ [$M_{\sun}$] &  $0.72 \pm 0.23$ &  This work$^{(e)}$, Sect. \ref{sec:RDM} \\ 
Corr(\RStar,\MStar) &   0.46    &  This work, Sect. \ref{sec:RDM} \\
 \hline
 \multicolumn{3}{c}{\textit{Stellar model inferences with C2kSMO}} \\
 \hline
$R_\mathrm{\star}$ [$R_{\sun}$]  &      $0.727 \pm 0.017$       &  This work, Sect. \ref{sec:stellarmodels} \\
$M_\mathrm{\star}$ [$M_{\sun}$]  &      $0.755 \pm 0.040$       &  This work, Sect. \ref{sec:stellarmodels} \\
$\rho_\star$ [$\rho_{\sun}$]  & $ 1.96 \pm 0.22$        &  This work, Sect. \ref{sec:stellarmodels} \\
Age$_\star$ [Gyr] & $9.3$               &  This work, Sect. \ref{sec:stellarmodels}  \\
 \hline
\end{tabularx}

\tablefoot{$^{(a)}$From $\pi$. $^{(b)}$Computed from \cite{Claret2011} tables. $^{(c)}$Averaged from the values available in the SIMBAD database \citep{CDS}; see text for details. $^{(d)}$From $\rho_\star$ and \RStar . $^{(e)}$From $\log(g)$.}
\tablebib{$^{(1)}$\citet{GaiaCat2018}\,; $^{(2)}$\cite{Oja1993}.}
\label{tab:StellarParam}
\end{table}

\subsection{Radius, density, and mass}
\label{sec:RDM}

The stellar radius is generally derived using the distance and angular diameter as follows:\, $\thetaLD=2\RStar/d$. As for the mass, \cite{Crida2018,Crida2018RNAAS} showed the importance of using the correlation between the stellar mass and radius to reduce the possible solutions in the mass-radius plane. We took the same approach to derive \RStar and \MStar. 
The PDF of \RStar, called $f_{R_\star}$, can be expressed as a function of the PDF of the observables \thetaLD (angular diameter) and $\pi$ (parallax), called $f_{\theta}$ and $f_{\pi}$ respectively. This gives\,
\begin{equation}
f_{R_\star}(R) =  \frac{R_0}{R^2}\int_0^\infty t\, f_{\pi}\left(\frac{R_0\,t}{R}\right)f_\theta(t)\ \dd t \ ,
\label{eq:PDFRetM}
\end{equation}
where $R_0$ is a constant \citep[see][for the proof]{Crida2018}. Concerning \textit{Gaia} parallaxes, \cite{Stassun2018} have reported that an offset of $-82\pm33~\mu as$ is observed, while \cite{Lindegren2018} have provided $-30~\mu as$. In any case, these offsets are within the uncertainty of the parallax for HD\,219134 and do not impact significantly our results. As advised by \cite{Luri2018}, we only used the parallax and its error given in the \textit{Gaia} DR2 catalogue \citep{Gaia+2016,GaiaCat2018}, keeping in mind the possible offsets and that for such bright stars, there might still be unknown offsets that DR3 and DR4 will provide.

We found \RStar$=0.726\pm0.014$ \rhoSun, which is a lower value than that found by \citet[][$\RStar=0.778\pm0.005$~\RSun]{Boyajian2012b}. Our uncertainty on \RStar is clearly dominated by the uncertainty on the angular diameter because we took the parallax from $Gaia$ DR2, which is very precise ($0.06\%$).

\begin{figure}
\includegraphics[scale=0.5]{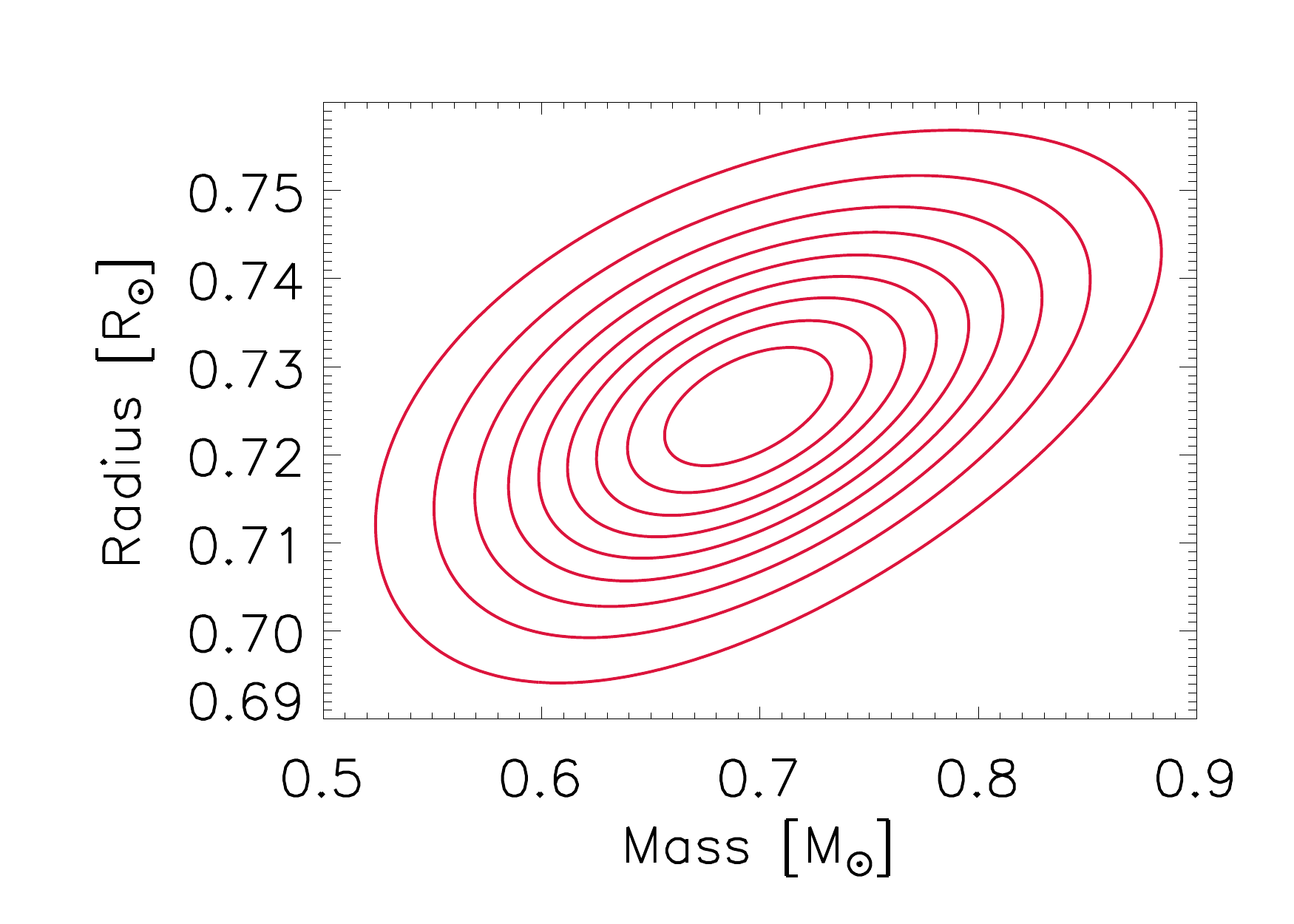}
\caption{Joint likelihood of the radius and mass of the star HD\,219134. The 9 plain red contour lines separate 10 equal-sized intervals between 0 and the maximum of Eq. (\ref{eq:MRstar_analytique}).}
\label{fig:PDFStar}
\end{figure}

The stellar density \rhoStar can be derived from the transit duration, period and depth \citep{Seager2003}. In our system, we have two transiting exoplanets. We computed the stellar density independently for both transits using the data given by \cite{Gillon2017} and found $1.74\pm0.22$ and $2.04\pm0.37$ \rhoSun for planets $b$ and $c,$ respectively. We note that the density coming from the analysis of the light curve for HD\,219134~$c$ is less precise than that of HD\,219134~$b$. This comes from the transit light curves themselves, which are more complete and more precise for planet $b$. Combining both densities, we obtained $1.82\pm0.19$ \rhoSun, which we use in the rest of our analysis. We computed the uncertainty following a classical propagation of errors and found a value close but different, and with a bigger error bar compared to that given in \cite{Gillon2017}. 
The joint likelihood of \MStar and \RStar can be expressed as\,
\begin{equation}
\Ld_{MR\star} (M,R) = \frac{3}{4\pi R^3}\times f_{R_\star}(R)\times f_{\rho_\star}\left(\frac{3M}{4\pi R^3}\right) \ ,
\label{eq:MRstar_analytique}
\end{equation}
as described in \cite{Crida2018} and where $f_{\rho_\star}$ is the PDF of the stellar density (Fig. \ref{fig:PDFStar}). The calculated correlation coefficient between \RStar and \MStar is 0.46. Our computation yields \MStar$=0.696\pm0.078$ \MSun, which is consistent with the value determined directly from \logg and \RStar but with a better precision. For reference, other authors derived $0.763\pm0.076 \MSun$ \citep{Boyajian2012b} using the relation by \cite{Henry1993}, and $0.81\pm0.03 \MSun$ \citep{Gillon2017} using stellar evolution modelling. In this latter case, the uncertainty corresponds to the internal source of error of the model and is thus underestimated.

\subsection{\textbf{Bolometric flux, effective temperature, and luminosity}}
\label{subsec:SED}

To derive the \Teff\ of the star we combined the angular diameter with its bolometric flux \Fbol\ using \begin{equation}
\Teff = \left( \frac{4 \times F_{\rm bol}} {\sigma_{\rm SB} \theta_{\rm LD}^2} \right)^{0.25} \ ,
\label{eq:Teff}
\end{equation}
where $\sigma_{\rm SB}$ is the Stefan-Boltzmann constant. 
This implies the computation of the bolometric flux, which we derived from the stellar photometry as described in the next subsection.

\subsubsection{Bolometric flux}

We determine the bolometric flux \Fbol\ and its uncertainty in the following way.  We retrieved photometric data from the literature made available by the VizieR Photometry tool\footnote{\texttt{http://vizier.u-strasbg.fr/vizier/sed/}}. These photometry converted-to-flux measurements were fitted to the BaSeL empirical library of spectra \citep{Lejeune97}, using a non-linear least-squared minimisation algorithm (Levenberg-Marquardt). The spectra are characterised by
\Teff, [M/H], and \logg. To convert these spectra to observed spectra they need to be scaled by ($R_{\star}/d$) and reddened for interstellar extinction A$_{\rm v}$.  Thus each model spectrum is characterised by these five parameters.   In practice most of these parameters are degenerate, so it is necessary to fix a subset of these.  
For each minimisation performed we fixed [M/H], \logg, and A$_{\rm v}$ i.e. we only fitted \Teff\ and ($R/d \propto \theta$), and then we integrated under the resulting scaled and unreddened empirical spectrum to obtain \Fbol. 

To properly estimate the uncertainties in the parameters we repeated this method 1000 times to obtain a distribution of \Fbol.  Each of these minimisations had different fixed values of [M/H], \logg,\ and A$_{\rm v}$ obtained by drawing random numbers from gaussian distributions characterised by the following: [M/H] = +0.07 $\pm$ 0.10, \logg\ = 4.57 $\pm$ 0.14, and A$_{\rm v} = 0.00 \pm 0.01$ mag, as discussed in Sect.~\ref{subsec:angdiam}. The initial values of \Teff\ and ($R/d$) were obtained by drawing them from a random uniform distribution with values between $4\,100$ and $5\,700$ K and between 4.0 and 8.0.   
Using the resulting distribution of \Fbol, we calculated \Fbol $= (19.86 \pm 0.21)\cdot10^{-8}$ erg s$^{-1}$ cm$^{-2}$.  In the same way, we also estimated the \Teff\ from the resulting distributions of the best-fitted \Teff\  ($T_{\rm eff, SED}$) and the angular diameter ($R/d$) converted to units of mas ($\theta_{\rm SED}$), although these latter two are not used any further in this work. 

\begin{figure}
    \centering
    \includegraphics[scale=0.3]{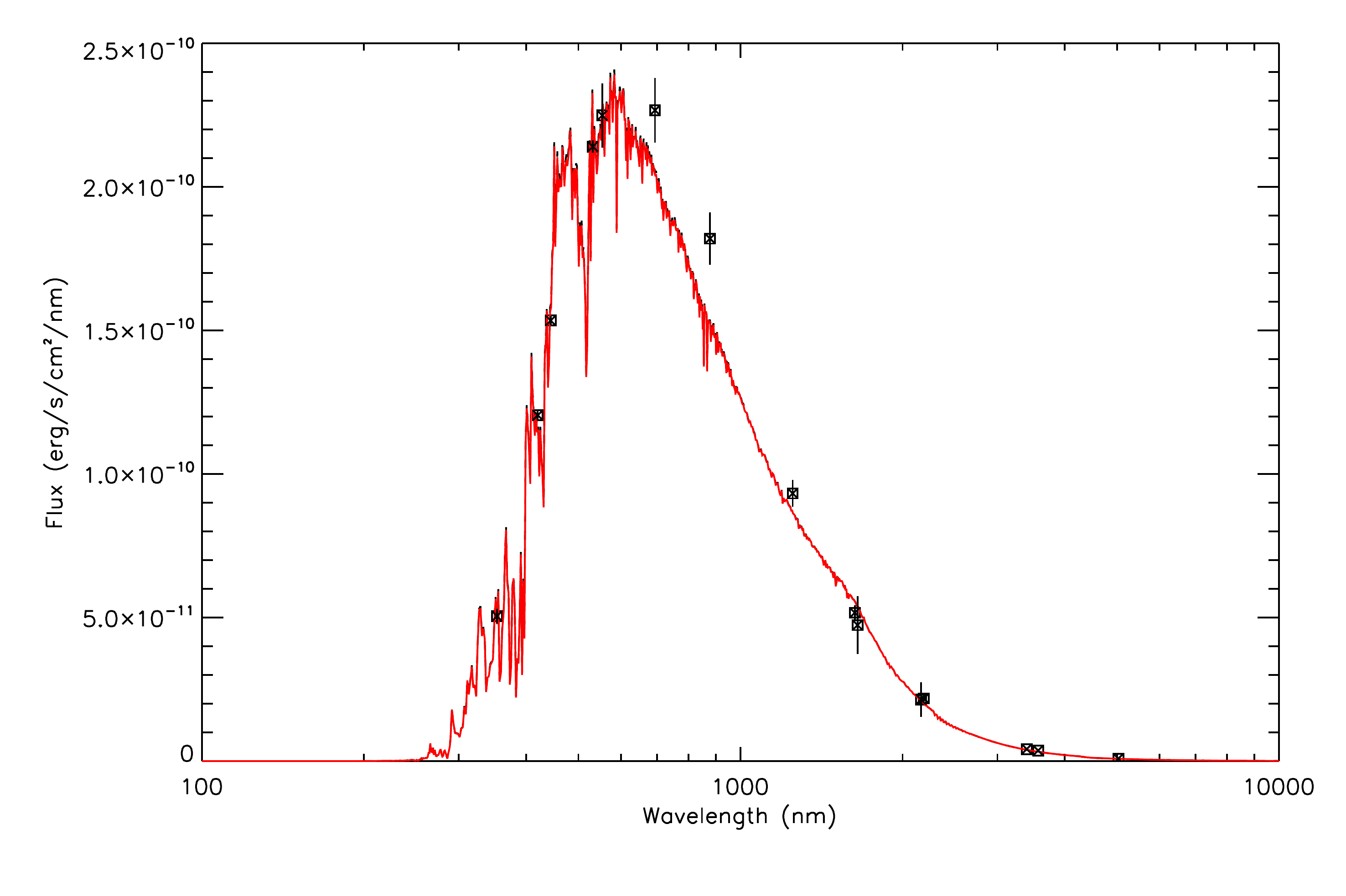}
    \caption{Photometric data (black squares) and fitted model (solid red line) from the BaSeL library of spectra.}
    \label{fig:fluxfit}
\end{figure}

The best-fit model spectrum is shown in Fig.~\ref{fig:fluxfit} in red, along with photometric data points in black.  
Overall, the model fits the data well, except for two points that are above the fit, but removing the two outliers did not change the results. These over-fluxes could come from a close star or another undetermined source, although we could not verify these hypotheses.

\subsubsection{Effective temperature and luminosity}

We derived the effective temperature \Teff from \Fbol and \thetaLD using Eq.~(\ref{eq:Teff}) to obtain $4\,858 \pm 50$ K. 
This is in very good agreement with the \Teff\ determined by Gaia ($4\,787 ^{+92}_{-73}$~K) and with that determined through the SED fitting ($4\,839 \pm 25$~K), which has a lower uncertainty. 
We finally obtained the luminosity using the distance and \Fbol\,as follows: 
\begin{equation}
L_{\star} = 4\pi d^2 F_{\rm bol} \ .
\label{eq:Lum}
\end{equation}
The errors on these final parameters were estimated following a classical propagation of errors \citep[see][for details]{Ligi2016}. The \textit{Gaia} luminosity is $L_\star = 0.30 L_{\sun}$, which is in good agreement with our value ($L_\star = 0.264 \pm 0.004 L_{\sun}$) considering the documented possible systematic errors. All final stellar parameters are reported in Table~\ref{tab:StellarParam}.

\subsection{Comparison with stellar evolution models}
\label{sec:stellarmodels}

HD\,219134 is now a well-characterised star thanks to our direct measurements of its radius and density, providing in turn its mass. Therefore, it constitutes a good benchmark to be compared to stellar evolution models. 

We thus confront our measurements (mass, radius, and density) to the values that can be inferred from stellar evolution modelling.  For that purpose, we have used the {\small C2kSMO}\footnote{ \small{C2kSMO} stands for ”Cesam2k  Stellar  Model  Optimisation”} stellar model optimisation pipeline \citep{Lebreton2014} to find the mass, age, and initial metallicity of the stellar model that best fits the luminosity, effective temperature, and surface metallicity (hereafter observational constraints) of HD\,219134 given in Table \ref{tab:StellarParam}. The procedure operates via a Levenberg-Marquardt minimisation performed on stellar models calculated on-the-fly with  the  Cesam2k  \citep{Morel2008} stellar evolution code \citep[\small{C2kSMO} is described in detail in ][]{Lebreton2014}.

For a given set of input parameters and physics of a stellar model  (nuclear reaction rates, equation of state, opacities, atmospheric boundary conditions, convection formalism, and related mixing-length parameter for convection, element diffusion and mixing, solar mixture of heavy elements, initial helium content, etc.), we can therefore infer the mass, age, and initial metallicity of the best model for HD\,219134 with internal error bars resulting from the uncertainties on the observational constraints.  
However, among these inputs, many are still very uncertain or even unknown. Accordingly, to get a reasonable estimate of the accuracy of the results, we
performed several model optimisations, each of which correspond to a different set of input physics and parameters.  
We varied the following -- most uncertain -- inputs:
\begin{itemize}
        \item {\it Solar mixture.} We investigated the effects of using either the GN93 \citep{GN93}  or the AGSS09 \citep{Asplund2009} mixture. However, we point out that, although still widely used, the GN93 mixture is no longer valid. The AGSS09 mixture is based on carefully updated atomic data and on a 3D time-dependent hydrodynamic model of the solar atmosphere, while the GN93 mixture was inferred through a 1D model of the solar atmosphere. As discussed by for example \citet[][]{2009LRSP....6....2N}, the 3D model reproduces the observations of the solar atmosphere
remarkably well, while the 1D model atmosphere does not. The AGSS09 mixture should therefore be preferred.
        \item   {\it Convection description.}  We used either the classical mixing-length theory \citep[usually referred to as MLT;][]{MLT} or the Canuto, Goldmann, and Mazzitelli formalism \citep[usually referred to as CGM; ][]{CGM}.
        \item {\it External boundary conditions.}  We investigated the effects of using either the approximate Eddington's grey radiative $T-\tau$ law ($T$ is the temperature, $\tau$ the optical depth) or the more physical $T-\tau$ law extracted from Model Atmospheres in Radiative and Convective Scheme (MARCS) model atmospheres \citep{Gustafsson2008}. Although MARCS models are classical 1D model atmospheres in local thermodynamical equilibrium, they do include convection in the MLT formalism and use up-to-date atomic and molecular data \citep[see e.g.][]{Gustafsson2008}. Therefore, these models represent an important progress with respect to the grey law and should be preferred.
        \item {\it Initial helium abundance.} This quantity is not accessible through the analysis of stellar spectra because helium lines are not formed in the spectra of cool and tepid stars. It is a major source of uncertainty in stellar model calculation. In stellar models the initial helium abundance is generally estimated  from the $\Delta Y/\Delta Z$ galactic enrichment law\footnote{$\Delta Y/\Delta Z=(Y-Y_\mathrm{P})/Z$, where $Y_P$ is the primordial helium abundance in mass fraction, and $Y$ and $Z$ are the current helium and metallicity mass fractions, respectively.} to overcome this difficulty. 
Two different $\Delta Y/\Delta Z$ values are usually used: the value obtained from solar model calibration\footnote{In the solar model calibration process, the evolution of a $1\ M_\odot$ model is calculated up to the known solar age. Its initial helium content and mixing-length parameter are fixed by the constraint that at solar age, the model has reached the observed values of the solar radius, luminosity, and surface metallicity.} \citep[chosen for instance in the new Bag of Stellar Tracks and Isochrones (BaSTI) stellar model grids; see][]{Hidalgo2018}, which is $\approx 1$ or the so-called galactic value,  $\Delta Y/\Delta Z\approx 2$ \citep{2007MNRAS.382.1516C} adopted for instance in the Modules for Experiments in Stellar Astrophysics (MESA) grids by \citet[][]{Coelho2015}. The former depends on the input physics of the solar model while the latter is very uncertain \citep[see e.g.][]{2010A&A...518A..13G}. On the other hand, the initial helium content can be estimated by modelling stars with available asteroseismic observational constraints. This is the case of 66 stars in the \textit{Kepler} Legacy sample for which we obtained values of $\Delta Y/\Delta Z$ in the range $1-3$ with a mean of $(\Delta Y/\Delta Z)_\mathrm{seism} \approx 2.3$ with the C2kSMO pipeline \citep[see e.g. ][]{2017ApJ...835..173S}.
Since no strong justification of what would be the best choice can be given, we  investigated the impact of using the two values $\Delta Y/\Delta Z=1$ and $2$ because the latter is also close to the mean \textit{Kepler} Legacy asteroseismic value $(\Delta Y/\Delta Z)_\mathrm{seism}$, but keeping in mind this remains the main source of uncertainty in our results.
\end{itemize}

More details on the uncertainties of stellar model inputs and their consequences can be found in \citet{2014EAS....65...99L}. To avoid such sources of uncertainties, direct measurements of stellar parameters should be preferred when possible.

Depending on the stellar  model input physics and parameters, we obtained a large range of possible ages, between $\approx 0.2$ and $9.3$ Gyr with large error bars. The range of possible masses is between $0.755$ and $0.810\ M_\odot$. The internal error bar on the inferred mass for an optimised stellar model based on a given set of inputs physics and parameters due to the uncertainty on the observational constraints (luminosity, effective temperature, and metallicity) is $\approx \pm 0.04\ M_\odot$. This error bar appears to be small. Indeed, in the Levenberg-Marquardt minimisation the error bars on the free parameters are obtained as the diagonal coefficients of the inverse of the Hessian matrix and have been shown to be smaller than those provided with other minimisation techniques \citep[see e.g. ][]{2017ApJ...835..173S}. The inferred stellar radii are in the range $0.727-0.728\ R_\odot$ with an internal error bar of $\pm 0.017\ R_\odot$, while the mean densities are in the range $1.96-2.09$ ($\pm 0.22$) \rhoSun. We chose as reference model for the star that based on the most appropriate input physics as explained in the description above (AGSS09 solar mixture and boundary conditions from MARCS model atmospheres), and the galactic value $\Delta Y/\Delta Z=2$ derived by \citet{2007MNRAS.382.1516C}, which is also rather close to the \textit{Kepler} Legacy seismic mean value $(\Delta Y/\Delta Z)_\mathrm{seism}$. This particular model has $M_\star = 0.755\pm 0.040\ M_\odot$ and an age of $9.3$ Gyr. Although this mass estimate is higher than the mass we derived from interferometry and transit by $\sim8\%$, the interval of solutions is consistent with our uncertainties. Similarly, our radius and density are consistent with those derived from the model ($0.727\pm0.017$ \RSun, $1.96\pm0.22$ \rhoSun, respectively). We point out that pushing the $\Delta Y/\Delta Z$ value from $2.$ to $3.$ would induce a change of mass from $0.755$ to $0.719\ M_\odot$, i.e. closer to the interferometric measure, but with a change in age from $9.3$ to $13.8$ Gyr, i.e. the age of the Universe; in our opinion this indicates that $\Delta Y/\Delta Z$ values that are too high are not realistic for this star.

We point out that, as is well-known in particular in the case of low-mass stars, the ages of stars are very poorly estimated when only the H-R diagram parameters and metallicity are known because of degeneracies in the stellar models \citep[see e.g.][]{2014EAS....65...99L,Ligi2016}. Furthermore, other values of the classical stellar parameters of HD\,219134 have been reported in the literature. To see how these reported values can modify our results we optimised stellar models on the basis of the \cite{Folsom2018} results on $\Teff$ and $\rm [Fe/H]$ and on $L_\star$ inferred from the SIMBAD Hipparcos V-magnitude. We obtained a similar range of masses $0.76-0.79 M_\odot$, while the models systematically point towards higher ages $10.2-13.8$ Gyr, which is mainly due to the smaller $\Teff$ ($4\,756 \pm 86$ K) derived by \cite{Folsom2018}. It is also worth pointing out that, as noted by \cite{Johnson2016}, the very high ages inferred from stellar models commonly found in the literature for HD\,219134 seem to be in conflict with ages from activity which, although not very precise, span the range $\approx 3-9$ Gyr\footnote{We estimated this age range from the empirical relation relating the Ca{\textsc{II}} H \& K emission index $R^\prime_\mathrm{ HK}$ and age derived by \citet{Mamajek2008}, with the value of $R^\prime_\mathrm{ HK}$ measured by \cite{BoroSaikia2018}.}.

\section{Planetary parameters and composition of the transiting exoplanets}
\label{sec:PlanetaryParam}

The precise and accurate stellar parameters that we have determined allow us to infer the parameters of the transiting exoplanets of the system. It is then possible to derive their internal composition using an inference scheme, and to verify if they stand in a dynamical point of view.

\subsection{Radius, density, and mass of the two transiting exoplanets}
\label{sec:RDMplanets}

\begin{figure}
\hspace{-0.8cm}
\includegraphics[scale=0.55]{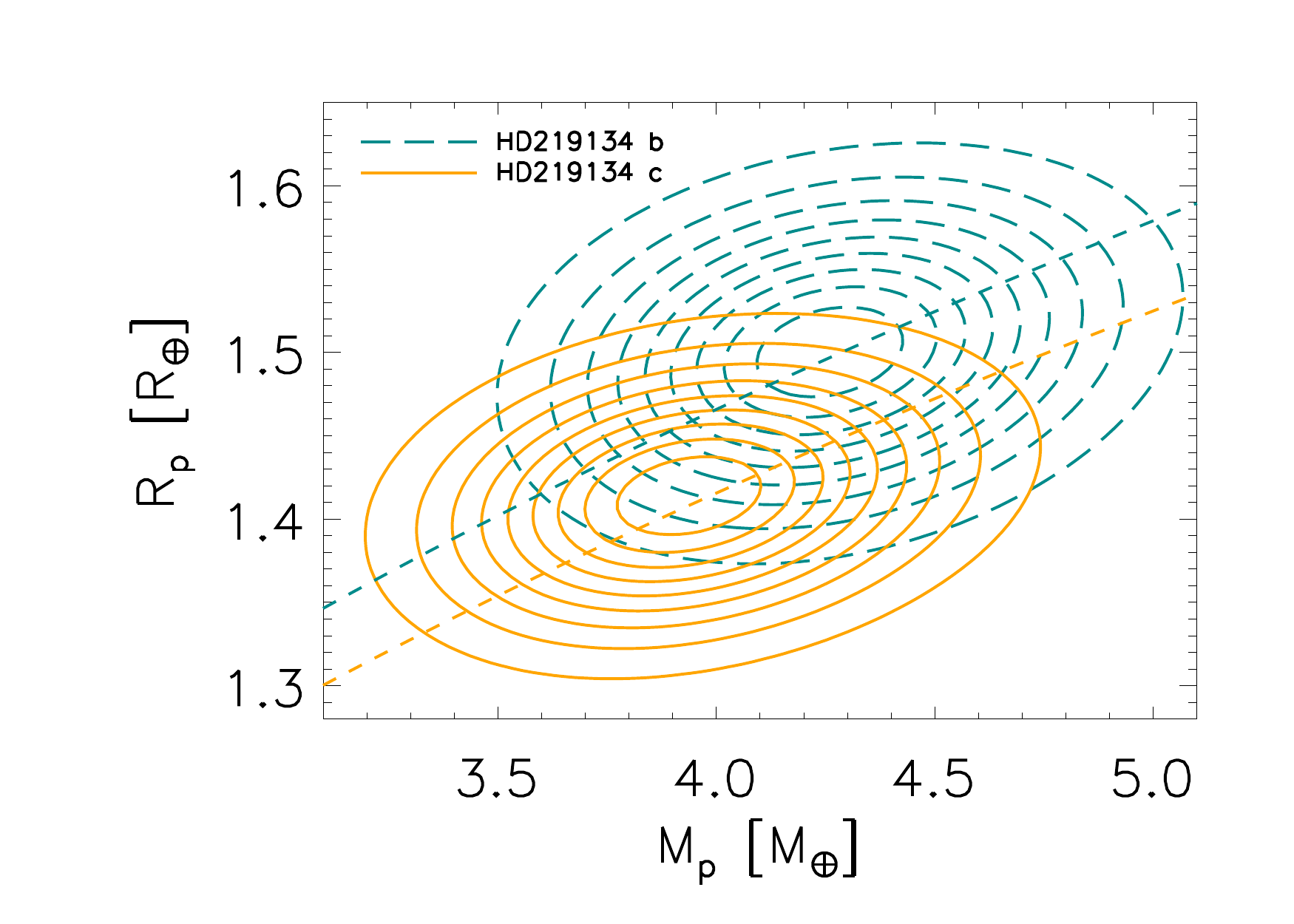}
\caption{Joint likelihood of the planetary mass and radius for planet $b$ (green long-dashed line) and planet $c$ (yellow solid line). The 9 contour lines separate 10 equal-sized intervals between 0 and the maximum of $f_p(M_p,R_p)$. The dashed lines show the iso-densities corresponding to the mean densities of planets $b$ and $c$.}
\label{fig:PDFplanets}
\end{figure}
The two planets HD\,219134~$b$ and $c$ transit their host star, and we can thus derive their properties. We computed the planetary radius \Rpl and mass \Mpl of each planet starting from the PDF of the stellar mass and radius. As explained by \cite{Crida2018} concerning 55\,Cnc~$e$, for any \Mpl and \MStar, we can derive the associated semi-amplitude of the RV signal $K$ following Kepler's law, and for any pair of \Rpl and \RStar, we can derive the associated transit depth \TD. We took the \TD,  $K$, and the period $P$ from \cite{Gillon2017} to calculate the PDF of the planetary mass and radius following the formula \cite[see Sect.~3.1 of][for more details]{Crida2018}\,:
\begin{equation}
\label{eq:PDF_MR_planet}
\begin{aligned}
f_p(M_p,R_p) \propto & \iint \exp\left(-\frac12\left(\frac{K(M_p,M_\star)-K}{\sigma_K}\right)^2\right) \\
 & \ \times \exp\left(-\frac12\left(\frac{\TD(M_p,M_\star)-\TD}{\sigma_{\TD}}\right)^2\right) \\
 & \ \times \Ld_{MR\star}(M_\star,R_\star)\  \dd M_\star\,\dd R_\star \ .
 \end{aligned}
\end{equation}
From this joint PDF, we compute the densities of both transiting exoplanets taking into account the correlation between \Rpl and \Mpl (Fig. \ref{fig:PDFplanets}).

\begin{table*}
\centering
\caption{Parameters of the innermost exoplanets of the system HD\,219134.}
\begin{tabular}{l | c | c | c | c }
\hline \hline
Param.                  &       HD\,219134 $b$                          & HD\,219134 $c$  & HD\,219134 $f$  & HD\,219134 $d$ \\
\hline
\Rpl    [\REarth]       &       $1.500 \pm 0.057$       &       $1.415 \pm 0.049$  &    -       &      -     \\
\Mpl    [\MEarth]       &       $4.27 \pm 0.34$ &       $3.96 \pm 0.34  $&    6.60      &      14.64     \\
Corr(\Rpl,\Mpl)                 &       0.22            &           0.23 &    -       &      -     \\
\rhopl [\rhoEarth]      &       $1.27 \pm 0.16$ &       $ 1.41 \pm 0.17$        &    -       &      -     \\
$a$ [au]              &  0.037 &  0.062 &  0.139  & 0.225 \\
\hline
\end{tabular}
\tablefoot{For the two transiting exoplanets, the mass is given, while for the two other planets the minimum planetary mass $\Mpl\sin(i)$ is given.}
\label{tab:ExoplParam}
\end{table*}

The new values of the planetary parameters are given in Table~\ref{tab:ExoplParam}. The radii of planets $b$ and $c$ are $1.500 \pm 0.057$ and $1.415 \pm 0.049$ \REarth, respectively. Because we find that the star is smaller than initially thought, the two planets appear smaller as well; \cite{Gillon2017} give \Rpl$=1.602\pm0.055$ and $1.511\pm0.047$ \REarth, and \Mpl$= 4.74\pm0.19$ and $4.36\pm0.22$ \MEarth, for planets $b$ and $c$, respectively. This enforces the idea that the two planets lie in the super-Earth part of the distribution of exoplanetary radii set by \cite{Fulton+2017}.

Even more interestingly, planet $c$ presents a higher density than planet $b,$ whereas it has smaller mass and radius. 
From the values in Table~\ref{tab:ExoplParam}, we get $\rho_b/\rho_c = 0.901 \pm 0.157$ assuming $\rho_b$ and $\rho_c$ to be independent variables. But $\rho_b$ and $\rho_c$ are slightly correlated as they both depend on the stellar parameters. Estimating directly the ratio, the stellar parameters simplify out\,to\begin{equation}
\frac{\rho_b}{\rho_c}=\frac{M_b/R_b^3}{M_c/R_c^3}=\left(\frac{P_b}{P_c}\right)^{1/3}\left(\frac{\TD_c}{\TD_b}\right)^{3/2}\left(\frac{K_b}{K_c}\right) = 0.905\pm0.131 \ ,
\label{eq:rho_ratio}
\end{equation}
where $P_b$ and $P_c$ are the orbital periods of the planets; we used a standard propagation of error. This is a larger difference than between the Earth and Venus (whose density is $0.944\,\rhoEarth$). 
A better knowledge of the transit depth would help discriminate between the density ratio and unity.
We investigated the causes of this potential disparity in the next section.

We also updated the values of the minimum masses of planets $f$ and $d$, which as expected we find lower than previous estimates, and of their semi-major axes (Table \ref{tab:ExoplParam}) using \cite{Gillon2017} orbital solutions, as these planets are confirmed by several independent detection. Finally, we determined the habitable zone (HZ) of the star to verify if any of the exoplanets of this system lie in this zone. To compute the HZ, we used the method described by \cite{Jones2006}, who adopted a conservative approach of this range of distances. We first computed the critical flux which depends on the \Teff of the star, and we derived the inner and outer boundaries of the HZ \citep[see Eq. (1a) to (2b) of][for details]{Jones2006}. As a result, we find that the HZ spreads from 0.46 to 0.91 au from the star and that no planet in the system is located in this area.

\setlength{\tabcolsep}{6pt}
\begin{table}
\caption{Median stellar abundances of HD~219134 from Hypatia catalog \citep{hinkel2014stellar} after the outliers and duplicate studies were removed. 
\label{tab:starabdata}}
\begin{center}
\begin{tabular}{ll}
\hline\noalign{\smallskip}
Parameter & HD~219134 \\
\hline\noalign{\smallskip}
$[\rm Fe/H]$ & $0.13 \pm 0.08$\\
$[\rm Mg/H]$ & $0.16 \pm 0.14$\\
$[\rm Si/H]$& $0.17\pm 0.15$\\
$[\rm Na/H]$& $0.22 \pm 0.08$\\
$[\rm Al/H]$& $0.26\pm 0.07$\\
$[\rm Ca/H]$& $0.13\pm 0.13$\\
\hline
\end{tabular}
\tablefoot{The unit is dex.}
\end{center}
\end{table}

\subsection{Internal compositions}
\label{sec:InternalComposition}

The new mass and radius estimates allowed us to investigate the planetary interiors. Interestingly, there is a 10\% density difference between the two planets (see Eq. (\ref{eq:rho_ratio})),  which are otherwise very similar in mass.
Although the uncertainty in the density ratio allows for no interior difference between the planets, it is worth investigating what could cause this possible difference, which is larger than that between the Earth and Venus with a probability of about 70\% (or stated differently, $30\%$ chance for a difference of less than 5$\%$).
In fact, Eq.~(\ref{eq:rho_ratio}) suggests that there is a 50\% chance that planet $b$ is more than 10\% less dense than planet $c$.

The lower density of planet $b$ can be associated with secondary atmospheres or a rock composition that is enriched in very refractory elements \citep{dorn2018new,dorn2018secondary}. Recently, \citet{bower2019linking} demonstrated that fully or partially molten mantle material can lower the bulk density of super-Earth up to 13\%. Therefore, a difference between the planetary densities may also be due to different melt fractions in both planets. In Sect.\ref{sec:InternalComposition}, we investigate this additional scenario and discuss its implications.

We start by solving an inference problem, for which we use the data of mass, radius (Sect. \ref{sec:RDMplanets}), stellar irradiation, and stellar abundances (Table \ref{tab:starabdata}) to infer the possible structures and compositions of both planets. Stellar abundances of rock-forming elements (e.g. Fe, Mg, Si) are used as proxies for the rocky interiors to reduce interior degeneracy as proposed by \citet{dorn2015can}. 
The differences between both planet interiors may provide evidence of their different formation or evolution history. 

\subsubsection{Inference scheme}

We used the inference scheme of \citet{dorn2017generalized}, which calculates possible interiors and their confidence ranges.
Our assumptions for the interior model are similar to those in \citet{dorn2017generalized} and are summarised in the following. Since these two planets are smaller than $\sim 1.8\REarth$, which is suggested to be the boundary between super-Earths and mini-Neptunes \citep{Fulton+2017}, we consider that the planets are made of iron-rich cores, silicate mantles, and terrestrial-type atmospheres. In addition to following \citet{dorn2017generalized}, we also allowed for some reduction of the mantle density as caused by a high melt fraction.

The interior parameters comprise\begin{itemize}
\item Core size $r_{\rm core}$
\item Size of rocky interior $r_{\rm core+mantle}$\item Mantle composition (i.e. Fe/Si$_{\rm mantle}$, Mg/Si$_{\rm mantle}$)
\item Reduction factor of mantle density $f_{\rm mantle}$
\item Pressure imposed by gas envelope $P_{\rm env}$
\item Temperature of gas envelope parametrised by $\alpha$ (see Eq. (\ref{Tequa}))
\item Mean molecular weight of gas envelope $\mu$.\end{itemize}

The prior distributions of the interior parameters used in this study are stated in Table \ref{tab:priorinterior}.

\begin{table*}
\caption{Prior ranges for interior parameters.
 \label{tab:priorinterior}}
\begin{center}
\begin{tabular}{lll}
\hline
\hline\noalign{\smallskip}
Parameter & Prior range &  Distribution  \\
\noalign{\smallskip}
\hline
\hline\noalign{\smallskip}
Core radius $r_{\rm core}$         & (0.01  -- 1) $r_{\rm core+mantle}$ &uniform in $r_{\rm core}^3$\\
Fe/Si$_{\rm mantle}$        & 0 -- Fe/Si$_{\rm star}$&uniform\\
Mg/Si$_{\rm mantle}$      & Mg/Si$_{\rm star}$ &Gaussian\\
$f_{\rm mantle}$& 0. -- 0.2 & uniform\\
Size of rocky interior $r_{\rm core+mantle}$   & (0.01 -- 1) \Rpl& Uniform in $r_{\rm core+mantle}^3$\\
Pressure imposed by gas envelope $P_{\rm env}$           & 20 mbar -- 100 bar  &uniform in log-scale\\
Temperature of gas envelope $\alpha$             & $0.5-1$ & uniform \\
Mean molecular weight of gas envelope $\mu$             & 16 -- 50 g/mol & uniform\\
\hline
\end{tabular} 
\end{center}
\end{table*}

Our interior model uses a self-consistent thermodynamic model for solid state interiors from \citet{dorn2017generalized}. For any given set of interior parameters, this model allows us to calculate the respective mass, radius, and bulk abundances and to compare them to the actual observed data. The thermodynamic model comprises the equation of state (EoS) of pure iron by \citet{bouchet} and of the light alloy FeSi by \citet{hakim2018new}, assuming 2.5\% of FeSi similar to Earth's core. For the silicate-mantle, we used the model by \citet{connolly2009} to compute equilibrium mineralogy and density profiles given the database of  \citet{stixrude2011thermodynamics}.
We allowed for a reduction of mantle densities as caused by the presence of melt. Unfortunately, the knowledge of EoS of melts is limited for pressures that occur in super-Earths \citep[e.g.][]{spaulding2012evidence,bolis2016decaying,wolf2018equation}. Therefore, we decided to use a very simplified approach in that we used a fudge factor $f_{\rm mantle}$ that reduces the mantle density $\rho_{\rm mantle}$ in each grid layer $i$ by $\rho_{\rm mantle,i}\times (1 - f_{\rm mantle}$).

For the gas layer, we used a simplified atmospheric model for a thin, isothermal atmosphere in hydrostatic equilibrium and ideal gas behaviour, which is calculated using the scale-height model \citep[model II in][]{dorn2017generalized}. The model parameters that parametrise the gas layer and that we aim to constrain are the pressure at the bottom of the gas layer $P_{\rm env}$, the mean molecular weight $\mu$, and the mean temperature (parametrised by $\alpha$, see below).
The thickness of the opaque gas layer $d_{\rm env}$ is given by

\begin{equation}
d_{\rm env}= H \ln \frac{P_{\rm env}}{P_{\rm out}} \ ,
\end{equation}
where the amount of opaque scale heights $H$ is determined by the ratio of $P_{\rm env}$ and $P_{\rm out}$. The quantity $P_{\rm out}$ is the pressure level at the optical photosphere for a transit geometry that we fix to 20 mbar \citep{fortney2007planetary}. We allowed a maximum pressure $P_{\rm env}$ equivalent to a Venus-like atmosphere (i.e. 100 bar).
The scale height $H$ is expressed by
\begin{equation}
H = \frac{T_{\rm env\ } R^{*}}{g_{\rm env\ } \mu } \ ,
\end{equation}
where $g_{\rm env}$ and $T_{\rm env}$ are gravity at the bottom of the atmosphere and mean atmospheric temperature, respectively. The quantity $R^{*}$ is the universal gas constant (8.3144598 J mol$^{-1}$ K$^{-1}$) and $\mu$ the mean molecular weight.
The mass of the atmosphere $m_{\rm env}$ is directly related to the pressure $P_{\rm env}$ as

\begin{equation}\label{massgas}
m_{\rm env}= 4\pi  P_{\rm env} \frac{(\Rpl - d_{\rm env})^2}{g_{\rm env}} \ , \end{equation}
where $\Rpl - d_{\rm env}$ is the radius at the bottom of the atmosphere.

The atmosphere's constant temperature is defined as
\begin{equation}
\label{Tequa}
T_{\rm env}  = \alpha \Teff \sqrt{\frac{\RStar}{2 a}} \ ,
\end{equation}
where $a$ is the semi-major axis. The factor $\alpha$ accounts for possible cooling and warming of the atmosphere and can vary between 0.5 and 1, which is equivalent to the observed range of albedos among solar system bodies (0.05 for asteroids up to 0.96 for Eris). The upper limit of 1 is verified against the estimated $\alpha_{\rm max}$ \citep[see Appendix A in][]{dorn2017generalized}, which takes possible greenhouse warming into account.

\subsubsection{Inference results}

\begin{figure*}
\centering
\includegraphics[scale=0.85]{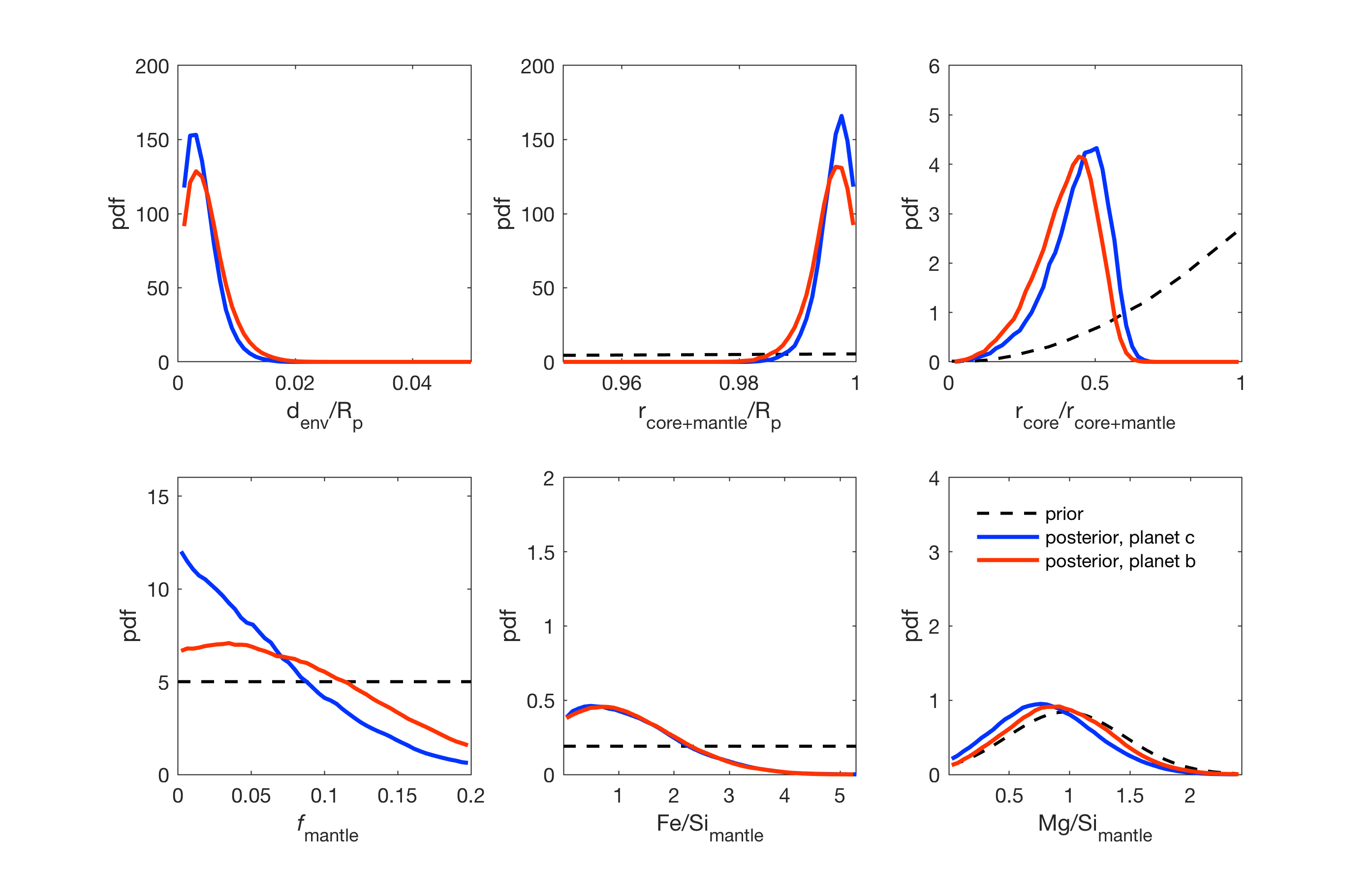}
\caption{One-dimensional marginalised posteriors of interior parameters: thickness of atmosphere ($d_{\rm env}$), size of rocky interior ($r_{\rm core+mantle}$), core size ($r_{\rm core}$), fudge factor $f_{\rm mantle}$, and mantle composition (Fe/Si$_{\rm mantle}$ and Mg/Si$_{\rm mantle}$). The prior distribution is shown in dashed lines (except for $d_{\rm env}$, for which no explicit prior is defined), while the posterior distribution is shown in solid lines for planets b (red) and c (blue).}
\label{fig:int}
\end{figure*}

Figure \ref{fig:int} summarises the interior estimates. Both planets have mantle compositions and core sizes that fit bulk density and the stellar abundance constraint. The core fraction of both planets is close to that of Venus and Earth ($(r_{\rm core}/r_{\rm core+mantle})_\oplus=0.53$), which validates their denomination as super-Earths. Compared to planet $c$, the lower density of 10\% of  planet
$b$ is associated with a slightly smaller core (by 10\%) and higher $f_{\rm mantle}$ (by 45\%), which indicates that a significantly stronger reduction of mantle density is plausible given the data. The estimates of $f_{\rm mantle}$ for planet $b$ and $c$ are $0.073_{-0.05}^{+0.06}$ and $0.05_{-0.04}^{+0.06}$, respectively.  Factors of $f_{\rm mantle}$ up to 0.25 can be associated with high melt fractions (for Earth-sized planets). Similar values can be achieved when the mantle composition is enriched by very refractory elements (i.e. Al, Ca). 

It should be noted that differences between the interiors are small, since uncertainties on bulk densities are relatively large. The data allow for no difference in bulk densities. However, a significant (more than 5\%) difference exists with 70\% probability. In this work, we used an interior model that allows us to quantify any possible difference in the rocky interiors of both planets. We assumed that any volatile layer is limited to a 100 bar atmosphere (similar to Venus) at maximum. Further arguments are necessary to evaluate whether a difference between the rocky interiors, specifically the mantle densities, can exist. 

Nonetheless, because \citet{bower2019linking} demonstrated that for Earth-sized planets a fully molten mantle is 25\% less dense than a solidified mantle, this possibility must be considered, and it is interesting to investigate whether planet $b$ could be less dense because partially molten.
Heating by irradiation from the host star would not be enough; the black-body equilibrium temperature for this planet is $1\,036$ K. Nevertheless, in the next subsection, we discuss a possible dynamical origin for the possible difference between HD\,219134 $b$ and $c$.

\subsection{Possible origin of a partial mantle melt for HD\,219134 $b$}
\label{sec:simus}

Large melt fractions may be sustained on planet $b$ by tidal heating. In the case of synchronous rotation with spin-orbit alignment, which is likely for close-in planets such as HD\,219134 $b$, tidal dissipation acts only on planets on eccentric orbits around the star.
The power is given by \citep[see e.g.][]{Lainey+2009}\,
\begin{equation}
\dot{E} = \frac{21}{2}\frac{k_2}{Q}\frac{(\omega R_p)^5}{G}e^2\ ,
\label{eq:Ptides}
\end{equation}
where $k_2$ is the Love number and $Q$ the quality factor of the planet of radius $R_p$ and spin or orbital frequency $\omega$. The key parameter $\frac{k_2}{Q}$ depends on the internal properties of the body\footnote{For reference, it is of the order of $10^{-4,-5}$ for gas giant planets and about $0.025$ for the Earth.}. 
The dissipated energy $\dot{E}$ heats the planet and damps the eccentricity of the orbit, ultimately leading to its circularisation and a reduction of the semi-major axis. To maintain tidal heating, the orbital eccentricity must be excited by the interaction with other secondary objects, as is the case for Jupiter's moon Io for instance. In order to investigate if tidal heating on planet $b$ is sufficient enough, we ran numerical simulations of the planetary system using the N-body code SyMBA \citep{Duncan+1998}.

To build our initial conditions, we took the $e$, $\varpi$, orbital periods, $K$, and mid-transit time from \cite{Gillon2017}.
They measure a non-zero eccentricity for planets $c$, $f$, and $d$, but not for planet $b$, whose eccentricity is fixed to zero to fit the other orbital parameters.
They do not provide data for the outermost two planets $g$ and $h$, but the long orbital periods of these planets make them unlikely to affect the inner four planets, and their orbital parameters suffer larger uncertainty so we neglect them in our simulations. We find that the eccentricity of planet $b$ is excited by the other planets. In absence of dissipation, the system is stable for at least 1 Gyr, and $e_b$ oscillates freely between $0$ and $0.13$ with a period of a few thousand years\footnote{Using initial circular orbits, i.e. assuming that the planets were fully formed locally in the protoplanetary disc, we observe no increase of the eccentricities of the four planets in 500 Myrs. This is not compatible with the observations of \citet{Gillon2017}; this suggests that these four planets may not have acquired their final mass and/or orbits during the protoplanetary disc phase. A phase of giant impacts or the breaking of a resonance chain \citep{Izidoro+2017,Pichierri+2018} could have happened in the early history of the system.}.

\begin{figure}
\hspace{-0.8cm}
\includegraphics{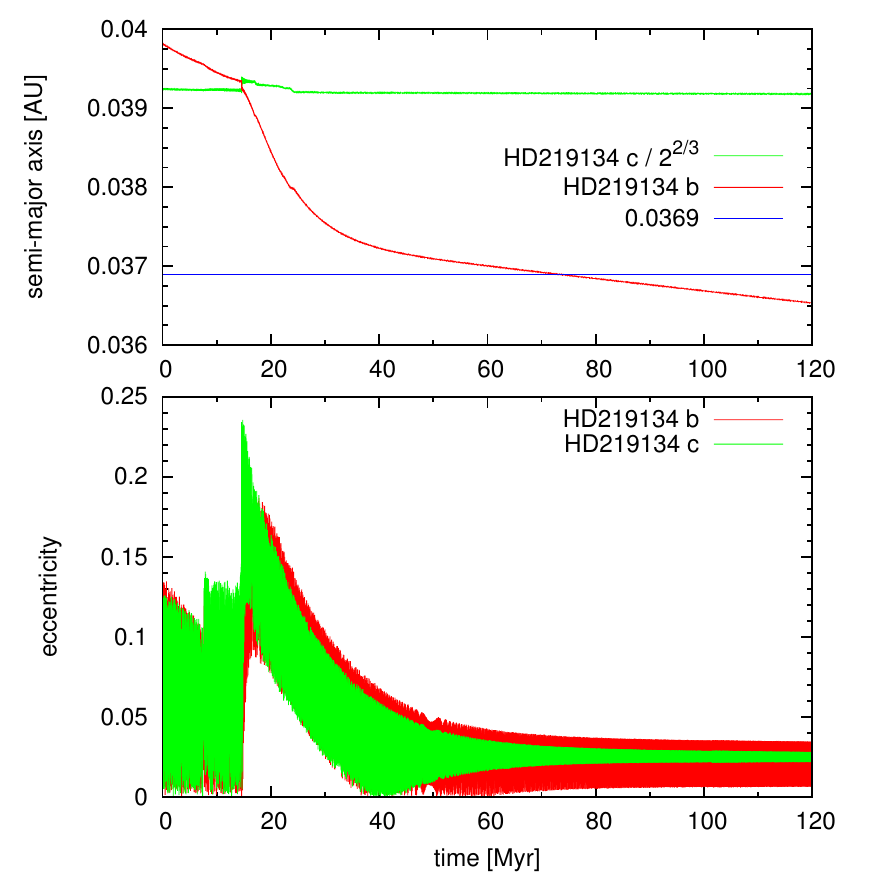}
\caption{Evolution of the inner two planets of the HD219134 system under dissipation with $k_2/Q=0.025$ for both planets, and in presence of planets $f$ and $d$ of \citet{Gillon2017}. Top panel: semi-major axis of planet $b$ (red curve), location of the 1:2 mean motion resonance with planet $c$ (green curve, that is $a_c/2^{2/3}$), and present semi-major axis of planet $b$ (blue horizontal line). Bottom panel: eccentricities of planets $b$ (red) and $c$ (green). }
\label{fig:Gillon}
\end{figure}

Introducing dissipation in planets $b$ and $c$, the eccentricities are damped and $e_b$ settles to a regime where it oscillates between $0.01$ and $0.06$. The energy loss is balanced by an inward drift of the planets, mainly planet $b$. We note that the final value of the eccentricity is independent of the assumed value for $k_2/Q$, only the timescale of the evolution and inward drift are proportional to $k_2/Q$. Because of this dissipation, the period ratio $P_c/P_b$ increases with time, and it is possible that this ratio (which is now $2.19$) was smaller than $2$, so that the 2:1 mean motion resonance was crossed recently. To check the effect of this phenomenon, we start planets $c$ and especially $b$ slightly out of their present position, inside the 2:1 mean motion resonance. Crossing the resonance at 14.6 Myrs kicks the eccentricities of planets $b$ and $c$, but this is quickly damped and the eccentricity of planet $b$ ends up oscillating between $0.005$ and $0.037$ with a period of $\sim 3000$ years when it reaches its present semi-major axis at 73 Myrs, as shown in Fig.~\ref{fig:Gillon}. Meanwhile, $e_c$ converges to $0.025$ \citep[while][find $0.062 \pm 0.039$]{Gillon2017}. We checked that again, $k_2/Q$ has little influence on the final behaviour of the eccentricities, although the speed at which the resonance is crossed matters.

Using Eq.~(\ref{eq:Ptides}), $0.005<e_b<0.037$ gives a total power for the tidal heating oscillating in $5.6 - 308 \times 10^{16}\times\left(\frac{k_2/Q}{0.025}\right)$~W for planet $b$ and around $2.1\times 10^{16}\times\left(\frac{k_2/Q}{0.025}\right)$~W for planet $c$.
For reference, tidal heating in Io is of the order of $10^{14}$~W \citep{Lainey+2009} so that, assuming $k_2/Q=0.025$ like for Earth, planet $b$ receives at least 2 and up to 100 times more tidal heating per mass unit than Io (and almost 300 with $e_b=0.06$). 
In contrast, because in Eq.~(\ref{eq:Ptides}) the term $(\omega R_p)^5$ is 70 times smaller for planet $c$ than for planet $b$, all other parameters being equal, it should be heated much less. We find that it gets a bit less tidal heating than Io per mass unit, so it is unlikely to melt even partially. In the end, the idea of a partial (if not total) melt of the mantle of HD219134\,$b$ to explain its possibly lower density than planet $c$ is strongly supported by dynamics.
A refinement of the parameters of the system and a complete stability analysis would help but are beyond the scope of this paper.

\section{Summary and conclusions}
\label{sec:Conclusion}

We present a new analysis of the exoplanetary system HD\,219134. We observed the star with the VEGA/CHARA interferometer and measured an angular diameter of $1.035 \pm 0.021$ mas and a radius of $0.726 \pm 0.014$ \RSun. This radius is not significantly affected by the \textit{Gaia} offset, but new values from the DR3 or DR4 will allow us to refine \RStar. 
We used the transit parameters from \cite{Gillon2017} to measure the stellar density ($1.82\pm0.19$ $\rho_{\sun}$) directly, and we directly derived from these two measurements the stellar mass ($0.696\pm0.078 M_{\sun}$) and the correlation between \MStar and \RStar ($0.46$). 
We compare our parameters with those obtained with C2kSMO and find that the range of masses is compatible with the directly measured mass, although the best model gives a mass $8\%$ higher than the directly measured mass. This corresponds to an age of 9.3 Gyr, but a large range of ages is  possible (0.2-9.3 Gyr). Similarly, previous indirect determinations of \MStar show higher values than our measurement \citep[see e.g.][]{Boyajian2012b, Gillon2017}, but it has to be noted that they are based on a larger \RStar.

The system includes two transiting exoplanets, HD\,219134 $b$ and $c$, for which we reassess the parameters. Using our new \RStar and \MStar, we computed the PDF of the planetary masses and radii, which we find lower than previous estimates (since previous stellar parameters were higher), and the correlations between \Mpl and \Rpl. 
These new values clearly validate the super-Earth nature of the two planets by putting them out of the gap in the exoplanetary radii distribution noticed by \cite{Fulton+2017}. 
We could thus derive the densities of the planets, which appear to differ by $10\%$, although these values are possibly identical within the error bars (70\% chance that the difference is more than $5\%$). More interestingly, planet $b$ has a lower density than planet $c$ despite its higher mass. 
Using \cite{dorn2017generalized} inference scheme, we show that this difference in density can be attributed to a slightly smaller core and/or a significantly lower mantle density. The latter might be due to a molten fraction. Tidal heating might be the cause of such a melting, as we investigated using the SyMBA N-body code. Excited by the other planets, the eccentricity of planets $b$ and $c$ reaches $\sim 0.02$ with tidal dissipation. This could lead to considerable heating for planet $b$ (100 times more than on Io per mass unit, possibly leading to partial melting of the mantle), while planet $c$ is too far from the star for tidal heating to be more intense than on Io.
Hence, despite their possible density difference, planets $b$ and $c$ may have the same composition, as expected in all standard planet formation models.

The system of HD\,219134 constitutes a benchmark case for both stellar and planetary sciences. Our direct estimation of the stellar radius and mass directly impacts the planetary parameters. Although within the error bars of the mass coming from C2kSMO, our new mass changes the planetary mass and the possibilities of interior structures compared to the possible solutions using the stellar models. 
Improving the precision of the transit light curves of the two planets would allow us to reduce the uncertainty on the stellar density, hence on the stellar mass. It would reduce the uncertainty on the planetary parameters even more, potentially answering the question of the density ratio of the two transiting super-Earths.
More generally, measuring the stellar radius and density as we have done in this work is the most direct method to infer stellar (hence planetary) parameters and should be more extensively used; this approach will certainly be possible within the Transiting Exoplanet Survey Satellite (TESS) and PLAnetary Transits and Oscillations of stars (PLATO) missions era.

\begin{acknowledgements}
We thank the anonymous referee who brought useful comments and significantly helped improve our manuscript.
R.L. has received funding from the European Union's Horizon 2020 research and innovation programme under the Marie Sk\l odowska-Curie grant agreement n. 664931. C.D. acknowledges support from the Swiss National Science Foundation under grant PZ00P2\_174028.
We thank A. Nakajima for her help with the initial conditions of the N-body simulations.
This work is based upon observations obtained with the Georgia State University Center for High Angular Resolution Astronomy Array at Mount Wilson Observatory. The CHARA Array is supported by the National Science Foundation under Grants No. AST-1211929 and AST-1411654.
This research has made use of the SIMBAD database,
operated at CDS, Strasbourg, France.
This research has made use of the Jean-Marie Mariotti Center \texttt{SearchCal} service \footnote{Available at \texttt{http://www.jmmc.fr/searchcal}}
co-developed by LAGRANGE and IPAG, and of CDS Astronomical Databases SIMBAD and VIZIER \footnote{Available at \texttt{http://cdsweb.u-strasbg.fr/}}.
This work has made use of data from the European Space Agency (ESA) mission
{\it Gaia} (\texttt{https://www.cosmos.esa.int/gaia}), processed by the {\it Gaia}
Data Processing and Analysis Consortium (DPAC,
\texttt{https://www.cosmos.esa.int/web/gaia/dpac/consortium}). Funding for the DPAC
has been provided by national institutions, in particular the institutions
participating in the {\it Gaia} Multilateral Agreement.
\end{acknowledgements}

\bibliographystyle{aa}
\bibliography{article}

\begin{thebibliography}{96}
\expandafter\ifx\csname natexlab\endcsname\relax\def\natexlab#1{#1}\fi

\bibitem[{{Allende Prieto} {et~al.}(2004){Allende Prieto}, {Barklem},
  {Lambert}, \& {Cunha}}]{Allende2004}
{Allende Prieto}, C., {Barklem}, P.~S., {Lambert}, D.~L., \& {Cunha}, K. 2004,
  \aap, 420, 183

\bibitem[{{Asplund} {et~al.}(2009){Asplund}, {Grevesse}, {Sauval}, \&
  {Scott}}]{Asplund2009}
{Asplund}, M., {Grevesse}, N., {Sauval}, A.~J., \& {Scott}, P. 2009, \araa, 47,
  481

\bibitem[{{Baglin}(2003)}]{Baglin2003}
{Baglin}, A. 2003, Advances in Space Research, 31, 345

\bibitem[{{Baines} {et~al.}(2010){Baines}, {D{\"o}llinger}, {Cusano},
  {Guenther}, {Hatzes}, {McAlister}, {ten Brummelaar}, {Turner}, {Sturmann},
  {Sturmann}, {Goldfinger}, {Farrington}, \& {Ridgway}}]{Baines2010}
{Baines}, E.~K., {D{\"o}llinger}, M.~P., {Cusano}, F., {et~al.} 2010, \apj,
  710, 1365

\bibitem[{{Boeche} \& {Grebel}(2016)}]{Boeche2016}
{Boeche}, C. \& {Grebel}, E.~K. 2016, \aap, 587, A2

\bibitem[{{B{\"o}hm-Vitense}(1958)}]{MLT}
{B{\"o}hm-Vitense}, E. 1958, \zap, 46, 108

\bibitem[{Bolis {et~al.}(2016)Bolis, Morard, Vinci, Ravasio, Bambrink,
  Guarguaglini, Koenig, Musella, Remus, Bouchet, {et~al.}}]{bolis2016decaying}
Bolis, R., Morard, G., Vinci, T., {et~al.} 2016, Geophysical Research Letters,
  43, 9475

\bibitem[{{Bonneau} {et~al.}(2006){Bonneau}, {Clausse}, {Delfosse}, {Mourard},
  {Cetre}, {Chelli}, {Cruzal{\`e}bes}, {Duvert}, \& {Zins}}]{Bonneau2006}
{Bonneau}, D., {Clausse}, J.-M., {Delfosse}, X., {et~al.} 2006, \aap, 456, 789

\bibitem[{{Boro Saikia} {et~al.}(2018){Boro Saikia}, {Marvin}, {Jeffers},
  {Reiners}, {Cameron}, {Marsden}, {Petit}, {Warnecke}, \&
  {Yadav}}]{BoroSaikia2018}
{Boro Saikia}, S., {Marvin}, C.~J., {Jeffers}, S.~V., {et~al.} 2018, \aap, 616,
  A108

\bibitem[{{Borucki} {et~al.}(2010){Borucki}, {Koch}, {Basri}, {Batalha},
  {Brown}, {Caldwell}, {Caldwell}, {Christensen-Dalsgaard}, {Cochran},
  {DeVore}, {Dunham}, {Dupree}, {Gautier}, {Geary}, {Gilliland}, {Gould},
  {Howell}, {Jenkins}, {Kondo}, {Latham}, {Marcy}, {Meibom}, {Kjeldsen},
  {Lissauer}, {Monet}, {Morrison}, {Sasselov}, {Tarter}, {Boss}, {Brownlee},
  {Owen}, {Buzasi}, {Charbonneau}, {Doyle}, {Fortney}, {Ford}, {Holman},
  {Seager}, {Steffen}, {Welsh}, {Rowe}, {Anderson}, {Buchhave}, {Ciardi},
  {Walkowicz}, {Sherry}, {Horch}, {Isaacson}, {Everett}, {Fischer}, {Torres},
  {Johnson}, {Endl}, {MacQueen}, {Bryson}, {Dotson}, {Haas}, {Kolodziejczak},
  {Van Cleve}, {Chandrasekaran}, {Twicken}, {Quintana}, {Clarke}, {Allen},
  {Li}, {Wu}, {Tenenbaum}, {Verner}, {Bruhweiler}, {Barnes}, \&
  {Prsa}}]{Borucky2010}
{Borucki}, W.~J., {Koch}, D., {Basri}, G., {et~al.} 2010, Science, 327, 977

\bibitem[{Bouchet {et~al.}(2013)Bouchet, Mazevet, Morard, Guyot, \&
  Musella}]{bouchet}
Bouchet, J., Mazevet, S., Morard, G., Guyot, F., \& Musella, R. 2013, Physical
  Review B, 87, 094102

\bibitem[{{Bourg{\'e}s} {et~al.}(2014){Bourg{\'e}s}, {Lafrasse}, {Mella},
  {Chesneau}, {Bouquin}, {Duvert}, {Chelli}, \& {Delfosse}}]{Bourges2014}
{Bourg{\'e}s}, L., {Lafrasse}, S., {Mella}, G., {et~al.} 2014, in Astronomical
  Society of the Pacific Conference Series, Vol. 485, Astronomical Data
  Analysis Software and Systems XXIII, ed. N.~{Manset} \& P.~{Forshay}, 223

\bibitem[{{Bourrier} {et~al.}(2018){Bourrier}, {Dumusque}, {Dorn}, {Henry},
  {Astudillo-Defru}, {Rey}, {Benneke}, {H{\'e}brard}, {Lovis}, {Demory},
  {Moutou}, \& {Ehrenreich}}]{Bourrier2018}
{Bourrier}, V., {Dumusque}, X., {Dorn}, C., {et~al.} 2018, \aap, 619, A1

\bibitem[{{Bower} {et~al.}(2019){Bower}, {Kitzmann}, {Wolf}, {Sanan}, {Dorn},
  \& {Oza}}]{bower2019linking}
{Bower}, D.~J., {Kitzmann}, D., {Wolf}, A.~S., {et~al.} 2019, arXiv e-prints,
  arXiv:1904.08300

\bibitem[{{Boyajian} {et~al.}(2012{\natexlab{a}}){Boyajian}, {McAlister}, {van
  Belle}, {Gies}, {ten Brummelaar}, {von Braun}, {Farrington}, {Goldfinger},
  {O'Brien}, {Parks}, {Richardson}, {Ridgway}, {Schaefer}, {Sturmann},
  {Sturmann}, {Touhami}, {Turner}, \& {White}}]{Boyajian2012a}
{Boyajian}, T.~S., {McAlister}, H.~A., {van Belle}, G., {et~al.}
  2012{\natexlab{a}}, \apj, 746, 101

\bibitem[{{Boyajian} {et~al.}(2012{\natexlab{b}}){Boyajian}, {von Braun}, {van
  Belle}, {McAlister}, {ten Brummelaar}, {Kane}, {Muirhead}, {Jones}, {White},
  {Schaefer}, {Ciardi}, {Henry}, {L{\'o}pez-Morales}, {Ridgway}, {Gies}, {Jao},
  {Rojas-Ayala}, {Parks}, {Sturmann}, {Sturmann}, {Turner}, {Farrington},
  {Goldfinger}, \& {Berger}}]{Boyajian2012b}
{Boyajian}, T.~S., {von Braun}, K., {van Belle}, G., {et~al.}
  2012{\natexlab{b}}, \apj, 757, 112

\bibitem[{{Canuto} {et~al.}(1996){Canuto}, {Goldman}, \& {Mazzitelli}}]{CGM}
{Canuto}, V.~M., {Goldman}, I., \& {Mazzitelli}, I. 1996, \apj, 473, 550

\bibitem[{{Casagrande} {et~al.}(2007){Casagrande}, {Flynn}, {Portinari},
  {Girardi}, \& {Jimenez}}]{2007MNRAS.382.1516C}
{Casagrande}, L., {Flynn}, C., {Portinari}, L., {Girardi}, L., \& {Jimenez}, R.
  2007, \mnras, 382, 1516

\bibitem[{{Chelli} {et~al.}(2016){Chelli}, {Duvert}, {Bourg{\`e}s}, {Mella},
  {Lafrasse}, {Bonneau}, \& {Chesneau}}]{Chelli2016}
{Chelli}, A., {Duvert}, G., {Bourg{\`e}s}, L., {et~al.} 2016, \aap, 589, A112

\bibitem[{{Claret} \& {Bloemen}(2011)}]{Claret2011}
{Claret}, A. \& {Bloemen}, S. 2011, \aap, 529, A75

\bibitem[{{Coelho} {et~al.}(2015){Coelho}, {Chaplin}, {Basu}, {Serenelli},
  {Miglio}, \& {Reese}}]{Coelho2015}
{Coelho}, H.~R., {Chaplin}, W.~J., {Basu}, S., {et~al.} 2015, \mnras, 451, 3011

\bibitem[{Connolly(2009)}]{connolly2009}
Connolly, J. 2009, Geochemistry, Geophysics, Geosystems, 10

\bibitem[{{Creevey} {et~al.}(2015){Creevey}, {Th{\'e}venin}, {Berio}, {Heiter},
  {von Braun}, {Mourard}, {Bigot}, {Boyajian}, {Kervella}, \&
  {Morel}}]{Creevey2015}
{Creevey}, O.~L., {Th{\'e}venin}, F., {Berio}, P., {et~al.} 2015, \aap, 575,
  A26

\bibitem[{{Creevey} {et~al.}(2012){Creevey}, {Th{\'e}venin}, {Boyajian},
  {Kervella}, {Chiavassa}, {Bigot}, {M{\'e}rand }, {Heiter}, {Morel}, \&
  {Pichon}}]{Creevey2012}
{Creevey}, O.~L., {Th{\'e}venin}, F., {Boyajian}, T.~S., {et~al.} 2012, \aap,
  545, A17

\bibitem[{{Crida} {et~al.}(2018{\natexlab{a}}){Crida}, {Ligi}, {Dorn}, {Borsa},
  \& {Lebreton}}]{Crida2018RNAAS}
{Crida}, A., {Ligi}, R., {Dorn}, C., {Borsa}, F., \& {Lebreton}, Y.
  2018{\natexlab{a}}, Research Notes of the American Astronomical Society, 2,
  172

\bibitem[{{Crida} {et~al.}(2018{\natexlab{b}}){Crida}, {Ligi}, {Dorn}, \&
  {Lebreton}}]{Crida2018}
{Crida}, A., {Ligi}, R., {Dorn}, C., \& {Lebreton}, Y. 2018{\natexlab{b}},
  \apj, 860, 122

\bibitem[{{da Silva} {et~al.}(2015){da Silva}, {Milone}, \&
  {Rocha-Pinto}}]{DaSilva2015}
{da Silva}, R., {Milone}, A. d.~C., \& {Rocha-Pinto}, H.~J. 2015, \aap, 580,
  A24

\bibitem[{Dorn {et~al.}(2018)Dorn, Harrison, Bonsor, \& Hands}]{dorn2018new}
Dorn, C., Harrison, J.~H., Bonsor, A., \& Hands, T.~O. 2018, Monthly Notices of
  the Royal Astronomical Society, 484, 712

\bibitem[{Dorn \& Heng(2018)}]{dorn2018secondary}
Dorn, C. \& Heng, K. 2018, The Astrophysical Journal, 853, 64

\bibitem[{Dorn {et~al.}(2015)Dorn, Khan, Heng, Connolly, Alibert, Benz, \&
  Tackley}]{dorn2015can}
Dorn, C., Khan, A., Heng, K., {et~al.} 2015, Astronomy \& Astrophysics, 577,
  A83

\bibitem[{Dorn {et~al.}(2017)Dorn, Venturini, Khan, Heng, Alibert, Helled,
  Rivoldini, \& Benz}]{dorn2017generalized}
Dorn, C., Venturini, J., Khan, A., {et~al.} 2017, Astronomy \& Astrophysics,
  597, A37

\bibitem[{{Duncan} {et~al.}(1998){Duncan}, {Levison}, \& {Lee}}]{Duncan+1998}
{Duncan}, M.~J., {Levison}, H.~F., \& {Lee}, M.~H. 1998, \aj, 116, 2067

\bibitem[{{Folsom} {et~al.}(2018){Folsom}, {Fossati}, {Wood}, {Sreejith},
  {Cubillos}, {Vidotto}, {Alecian}, {Girish}, {Lichtenegger}, {Murthy},
  {Petit}, \& {Valyavin}}]{Folsom2018}
{Folsom}, C.~P., {Fossati}, L., {Wood}, B.~E., {et~al.} 2018, \mnras, 481, 5286

\bibitem[{Fortney {et~al.}(2007)Fortney, Marley, \&
  Barnes}]{fortney2007planetary}
Fortney, J.~J., Marley, M.~S., \& Barnes, J.~W. 2007, The Astrophysical
  Journal, 659, 1661

\bibitem[{{Frasca} {et~al.}(2009){Frasca}, {Covino}, {Spezzi}, {Alcal{\'a}},
  {Marilli}, {F{\.z}r{\'e}sz}, \& {Gandolfi}}]{Frasca2009}
{Frasca}, A., {Covino}, E., {Spezzi}, L., {et~al.} 2009, \aap, 508, 1313

\bibitem[{{Fulton} {et~al.}(2017){Fulton}, {Petigura}, {Howard}, {Isaacson},
  {Marcy}, {Cargile}, {Hebb}, {Weiss}, {Johnson}, \& {Morton}}]{Fulton+2017}
{Fulton}, B.~J., {Petigura}, E.~A., {Howard}, A.~W., {et~al.} 2017, \aj, 154,
  109

\bibitem[{{Gaia Collaboration} {et~al.}(2018){Gaia Collaboration}, {Brown},
  {Vallenari}, {Prusti}, {de Bruijne}, {Babusiaux}, {Bailer-Jones}, {Biermann},
  {Evans}, {Eyer}, {Jansen}, {Jordi}, {Klioner}, {Lammers}, {Lindegren},
  {Luri}, {Mignard}, {Panem}, {Pourbaix}, {Randich}, {Sartoretti}, {Siddiqui},
  {Soubiran}, {van Leeuwen}, {Walton}, {Arenou}, {Bastian}, {Cropper},
  {Drimmel}, {Katz}, {Lattanzi}, {Bakker}, {Cacciari}, {Casta{\~n}eda},
  {Chaoul}, {Cheek}, {De Angeli}, {Fabricius}, {Guerra}, {Holl}, {Masana},
  {Messineo}, {Mowlavi}, {Nienartowicz}, {Panuzzo}, {Portell}, {Riello},
  {Seabroke}, {Tanga}, {Th{\'e}venin}, {Gracia-Abril}, {Comoretto},
  {Garcia-Reinaldos}, {Teyssier}, {Altmann}, {Andrae}, {Audard},
  {Bellas-Velidis}, {Benson}, {Berthier}, {Blomme}, {Burgess}, {Busso},
  {Carry}, {Cellino}, {Clementini}, {Clotet}, {Creevey}, {Davidson}, {De
  Ridder}, {Delchambre}, {Dell'Oro}, {Ducourant},
  {Fern{\'a}ndez-Hern{\'a}ndez}, {Fouesneau}, {Fr{\'e}mat}, {Galluccio},
  {Garc{\'\i}a-Torres}, {Gonz{\'a}lez-N{\'u}{\~n}ez}, {Gonz{\'a}lez-Vidal},
  {Gosset}, {Guy}, {Halbwachs}, {Hambly}, {Harrison}, {Hern{\'a}ndez},
  {Hestroffer}, {Hodgkin}, {Hutton}, {Jasniewicz}, {Jean-Antoine-Piccolo},
  {Jordan}, {Korn}, {Krone-Martins}, {Lanzafame}, {Lebzelter}, {L{\"o}ffler},
  {Manteiga}, {Marrese}, {Mart{\'\i}n-Fleitas}, {Moitinho}, {Mora}, {Muinonen},
  {Osinde}, {Pancino}, {Pauwels}, {Petit}, {Recio-Blanco}, {Richards},
  {Rimoldini}, {Robin}, {Sarro}, {Siopis}, {Smith}, {Sozzetti}, {S{\"u}veges},
  {Torra}, {van Reeven}, {Abbas}, {Abreu Aramburu}, {Accart}, {Aerts},
  {Altavilla}, {{\'A}lvarez}, {Alvarez}, {Alves}, {Anderson}, {Andrei},
  {Anglada Varela}, {Antiche}, {Antoja}, {Arcay}, {Astraatmadja}, {Bach},
  {Baker}, {Balaguer-N{\'u}{\~n}ez}, {Balm}, {Barache}, {Barata}, {Barbato},
  {Barblan}, {Barklem}, {Barrado}, {Barros}, {Barstow}, {Bartholom{\'e}
  Mu{\~n}oz}, {Bassilana}, {Becciani}, {Bellazzini}, {Berihuete}, {Bertone},
  {Bianchi}, {Bienaym{\'e}}, {Blanco-Cuaresma}, {Boch}, {Boeche}, {Bombrun},
  {Borrachero}, {Bossini}, {Bouquillon}, {Bourda}, {Bragaglia}, {Bramante},
  {Breddels}, {Bressan}, {Brouillet}, {Br{\"u}semeister}, {Brugaletta},
  {Bucciarelli}, {Burlacu}, {Busonero}, {Butkevich}, {Buzzi}, {Caffau},
  {Cancelliere}, {Cannizzaro}, {Cantat-Gaudin}, {Carballo}, {Carlucci},
  {Carrasco}, {Casamiquela}, {Castellani}, {Castro-Ginard}, {Charlot},
  {Chemin}, {Chiavassa}, {Cocozza}, {Costigan}, {Cowell}, {Crifo}, {Crosta},
  {Crowley}, {Cuypers}, {Dafonte}, {Damerdji}, {Dapergolas}, {David}, {David},
  {de Laverny}, {De Luise}, {De March}, {de Martino}, {de Souza}, {de Torres},
  {Debosscher}, {del Pozo}, {Delbo}, {Delgado}, {Delgado}, {Di Matteo},
  {Diakite}, {Diener}, {Distefano}, {Dolding}, {Drazinos}, {Dur{\'a}n},
  {Edvardsson}, {Enke}, {Eriksson}, {Esquej}, {Eynard Bontemps}, {Fabre},
  {Fabrizio}, {Faigler}, {Falc{\~a}o}, {Farr{\`a}s Casas}, {Federici},
  {Fedorets}, {Fernique}, {Figueras}, {Filippi}, {Findeisen}, {Fonti},
  {Fraile}, {Fraser}, {Fr{\'e}zouls}, {Gai}, {Galleti}, {Garabato},
  {Garc{\'\i}a-Sedano}, {Garofalo}, {Garralda}, {Gavel}, {Gavras}, {Gerssen},
  {Geyer}, {Giacobbe}, {Gilmore}, {Girona}, {Giuffrida}, {Glass}, {Gomes},
  {Granvik}, {Gueguen}, {Guerrier}, {Guiraud}, {Guti{\'e}rrez-S{\'a}nchez},
  {Haigron}, {Hatzidimitriou}, {Hauser}, {Haywood}, {Heiter}, {Helmi}, {Heu},
  {Hilger}, {Hobbs}, {Hofmann}, {Holland}, {Huckle}, {Hypki}, {Icardi},
  {Jan{\ss}en}, {Jevardat de Fombelle}, {Jonker}, {Juh{\'a}sz}, {Julbe},
  {Karampelas}, {Kewley}, {Klar}, {Kochoska}, {Kohley}, {Kolenberg},
  {Kontizas}, {Kontizas}, {Koposov}, {Kordopatis}, {Kostrzewa-Rutkowska},
  {Koubsky}, {Lambert}, {Lanza}, {Lasne}, {Lavigne}, {Le Fustec}, {Le
  Poncin-Lafitte}, {Lebreton}, {Leccia}, {Leclerc}, {Lecoeur-Taibi},
  {Lenhardt}, {Leroux}, {Liao}, {Licata}, {Lindstr{\o}m}, {Lister}, {Livanou},
  {Lobel}, {L{\'o}pez}, {Managau}, {Mann}, {Mantelet}, {Marchal}, {Marchant},
  {Marconi}, {Marinoni}, {Marschalk{\'o}}, {Marshall}, {Martino}, {Marton},
  {Mary}, {Massari}, {Matijevi{\v{c}}}, {Mazeh}, {McMillan}, {Messina},
  {Michalik}, {Millar}, {Molina}, {Molinaro}, {Moln{\'a}r}, {Montegriffo},
  {Mor}, {Morbidelli}, {Morel}, {Morris}, {Mulone}, {Muraveva}, {Musella},
  {Nelemans}, {Nicastro}, {Noval}, {O'Mullane}, {Ord{\'e}novic},
  {Ord{\'o}{\~n}ez-Blanco}, {Osborne}, {Pagani}, {Pagano}, {Pailler},
  {Palacin}, {Palaversa}, {Panahi}, {Pawlak}, {Piersimoni}, {Pineau}, {Plachy},
  {Plum}, {Poggio}, {Poujoulet}, {Pr{\v{s}}a}, {Pulone}, {Racero}, {Ragaini},
  {Rambaux}, {Ramos-Lerate}, {Regibo}, {Reyl{\'e}}, {Riclet}, {Ripepi}, {Riva},
  {Rivard}, {Rixon}, {Roegiers}, {Roelens}, {Romero-G{\'o}mez}, {Rowell},
  {Royer}, {Ruiz-Dern}, {Sadowski}, {Sagrist{\`a} Sell{\'e}s}, {Sahlmann},
  {Salgado}, {Salguero}, {Sanna}, {Santana-Ros}, {Sarasso}, {Savietto},
  {Schultheis}, {Sciacca}, {Segol}, {Segovia}, {S{\'e}gransan}, {Shih},
  {Siltala}, {Silva}, {Smart}, {Smith}, {Solano}, {Solitro}, {Sordo}, {Soria
  Nieto}, {Souchay}, {Spagna}, {Spoto}, {Stampa}, {Steele},
  {Steidelm{\"u}ller}, {Stephenson}, {Stoev}, {Suess}, {Surdej}, {Szabados},
  {Szegedi-Elek}, {Tapiador}, {Taris}, {Tauran}, {Taylor}, {Teixeira},
  {Terrett}, {Teyssand ier}, {Thuillot}, {Titarenko}, {Torra Clotet}, {Turon},
  {Ulla}, {Utrilla}, {Uzzi}, {Vaillant}, {Valentini}, {Valette}, {van Elteren},
  {Van Hemelryck}, {van Leeuwen}, {Vaschetto}, {Vecchiato}, {Veljanoski},
  {Viala}, {Vicente}, {Vogt}, {von Essen}, {Voss}, {Votruba}, {Voutsinas},
  {Walmsley}, {Weiler}, {Wertz}, {Wevers}, {Wyrzykowski}, {Yoldas},
  {{\v{Z}}erjal}, {Ziaeepour}, {Zorec}, {Zschocke}, {Zucker}, {Zurbach}, \&
  {Zwitter}}]{GaiaCat2018}
{Gaia Collaboration}, {Brown}, A.~G.~A., {Vallenari}, A., {et~al.} 2018, \aap,
  616, A1

\bibitem[{{Gaia Collaboration} {et~al.}(2016){Gaia Collaboration}, {Prusti},
  {de Bruijne}, {Brown}, {Vallenari}, {Babusiaux}, {Bailer-Jones}, {Bastian},
  {Biermann}, {Evans}, \& et~al.}]{Gaia+2016}
{Gaia Collaboration}, {Prusti}, T., {de Bruijne}, J.~H.~J., {et~al.} 2016,
  \aap, 595, A1

\bibitem[{{Gennaro} {et~al.}(2010){Gennaro}, {Prada Moroni}, \&
  {Degl'Innocenti}}]{2010A&A...518A..13G}
{Gennaro}, M., {Prada Moroni}, P.~G., \& {Degl'Innocenti}, S. 2010, \aap, 518,
  A13

\bibitem[{{Gillon} {et~al.}(2017){Gillon}, {Demory}, {Van Grootel}, {Motalebi},
  {Lovis}, {Cameron}, {Charbonneau}, {Latham}, {Molinari}, {Pepe},
  {S{\'e}gransan}, {Sasselov}, {Udry}, {Mayor}, {Micela}, {Piotto}, \&
  {Sozzetti}}]{Gillon2017}
{Gillon}, M., {Demory}, B.-O., {Van Grootel}, V., {et~al.} 2017, Nature
  Astronomy, 1, 0056

\bibitem[{{Gray} {et~al.}(2003){Gray}, {Corbally}, {Garrison}, {McFadden}, \&
  {Robinson}}]{Gray2003}
{Gray}, R.~O., {Corbally}, C.~J., {Garrison}, R.~F., {McFadden}, M.~T., \&
  {Robinson}, P.~E. 2003, \aj, 126, 2048

\bibitem[{{Grevesse} \& {Noels}(1993)}]{GN93}
{Grevesse}, N. \& {Noels}, A. 1993, in Origin and Evolution of the Elements,
  ed. N.~{Prantzos}, E.~{Vangioni-Flam}, \& M.~{Casse}, 15--25

\bibitem[{{Gustafsson} {et~al.}(2008){Gustafsson}, {Edvardsson}, {Eriksson},
  {J{\o}rgensen}, {Nordlund}, \& {Plez}}]{Gustafsson2008}
{Gustafsson}, B., {Edvardsson}, B., {Eriksson}, K., {et~al.} 2008, \aap, 486,
  951

\bibitem[{Hakim {et~al.}(2018)Hakim, Rivoldini, Van~Hoolst, Cottenier, Jaeken,
  Chust, \& Steinle-Neumann}]{hakim2018new}
Hakim, K., Rivoldini, A., Van~Hoolst, T., {et~al.} 2018, Icarus, 313, 61

\bibitem[{{Heiter} \& {Luck}(2003)}]{Heiter2003}
{Heiter}, U. \& {Luck}, R.~E. 2003, \aj, 126, 2015

\bibitem[{{Henry} \& {McCarthy}(1993)}]{Henry1993}
{Henry}, T.~J. \& {McCarthy}, Jr., D.~W. 1993, \aj, 106, 773

\bibitem[{{Hidalgo} {et~al.}(2018){Hidalgo}, {Pietrinferni}, {Cassisi},
  {Salaris}, {Mucciarelli}, {Savino}, {Aparicio}, {Silva Aguirre}, \&
  {Verma}}]{Hidalgo2018}
{Hidalgo}, S.~L., {Pietrinferni}, A., {Cassisi}, S., {et~al.} 2018, \apj, 856,
  125

\bibitem[{Hinkel {et~al.}(2014)Hinkel, Timmes, Young, Pagano, \&
  Turnbull}]{hinkel2014stellar}
Hinkel, N.~R., Timmes, F., Young, P.~A., Pagano, M.~D., \& Turnbull, M.~C.
  2014, The Astronomical Journal, 148, 54

\bibitem[{{Huber} {et~al.}(2012){Huber}, {Ireland}, {Bedding}, {Brand{\~a}o},
  {Piau}, {Maestro}, {White}, {Bruntt}, {Casagrande}, {Molenda-{\.Z}akowicz},
  {Silva Aguirre}, {Sousa}, {Barclay}, {Burke}, {Chaplin},
  {Christensen-Dalsgaard}, {Cunha}, {De Ridder}, {Farrington}, {Frasca},
  {Garc{\'{\i}}a}, {Gilliland}, {Goldfinger}, {Hekker}, {Kawaler}, {Kjeldsen},
  {McAlister}, {Metcalfe}, {Miglio}, {Monteiro}, {Pinsonneault}, {Schaefer},
  {Stello}, {Stumpe}, {Sturmann}, {Sturmann}, {ten Brummelaar}, {Thompson},
  {Turner}, \& {Uytterhoeven}}]{Huber2012}
{Huber}, D., {Ireland}, M.~J., {Bedding}, T.~R., {et~al.} 2012, \apj, 760, 32

\bibitem[{{Izidoro} {et~al.}(2017){Izidoro}, {Ogihara}, {Raymond},
  {Morbidelli}, {Pierens}, {Bitsch}, {Cossou}, \& {Hersant}}]{Izidoro+2017}
{Izidoro}, A., {Ogihara}, M., {Raymond}, S.~N., {et~al.} 2017, \mnras, 470,
  1750

\bibitem[{{Johnson} {et~al.}(2016){Johnson}, {Endl}, {Cochran}, {Meschiari},
  {Robertson}, {MacQueen}, {Brugamyer}, {Caldwell}, {Hatzes}, \&
  {Ram{\'\i}rez}}]{Johnson2016}
{Johnson}, M.~C., {Endl}, M., {Cochran}, W.~D., {et~al.} 2016, \apj, 821, 74

\bibitem[{{Jones} {et~al.}(2006){Jones}, {Sleep}, \& {Underwood}}]{Jones2006}
{Jones}, B.~W., {Sleep}, P.~N., \& {Underwood}, D.~R. 2006, \apj, 649, 1010

\bibitem[{{Lainey} {et~al.}(2009){Lainey}, {Arlot}, {Karatekin}, \& {van
  Hoolst}}]{Lainey+2009}
{Lainey}, V., {Arlot}, J.-E., {Karatekin}, {\"O}., \& {van Hoolst}, T. 2009,
  \nat, 459, 957

\bibitem[{{Lallement} {et~al.}(2014){Lallement}, {Vergely}, {Valette},
  {Puspitarini}, {Eyer}, \& {Casagrande}}]{Lallement2014}
{Lallement}, R., {Vergely}, J.-L., {Valette}, B., {et~al.} 2014, \aap, 561, A91

\bibitem[{{Lebreton} \& {Goupil}(2014)}]{Lebreton2014}
{Lebreton}, Y. \& {Goupil}, M.~J. 2014, \aap, 569, A21

\bibitem[{{Lebreton} {et~al.}(2014){Lebreton}, {Goupil}, \&
  {Montalb{\'a}n}}]{2014EAS....65...99L}
{Lebreton}, Y., {Goupil}, M.~J., \& {Montalb{\'a}n}, J. 2014, in EAS
  Publications Series, Vol.~65, EAS Publications Series, 99--176

\bibitem[{{Lee} {et~al.}(2011){Lee}, {Beers}, {Allende Prieto}, {Lai},
  {Rockosi}, {Morrison}, {Johnson}, {An}, {Sivarani}, \& {Yanny}}]{Lee2011}
{Lee}, Y.~S., {Beers}, T.~C., {Allende Prieto}, C., {et~al.} 2011, \aj, 141, 90

\bibitem[{{Lejeune} {et~al.}(1997){Lejeune}, {Cuisinier}, \&
  {Buser}}]{Lejeune97}
{Lejeune}, T., {Cuisinier}, F., \& {Buser}, R. 1997, VizieR Online Data
  Catalog, 412, 50229

\bibitem[{{Ligi} {et~al.}(2016){Ligi}, {Creevey}, {Mourard}, {Crida},
  {Lagrange}, {Nardetto}, {Perraut}, {Schultheis}, {Tallon-Bosc}, \& {ten
  Brummelaar}}]{Ligi2016}
{Ligi}, R., {Creevey}, O., {Mourard}, D., {et~al.} 2016, \aap, 586, A94

\bibitem[{{Ligi} {et~al.}(2012){Ligi}, {Mourard}, {Lagrange}, {Perraut},
  {Boyajian}, {B{\'e}rio}, {Nardetto}, {Tallon-Bosc}, {McAlister}, {ten
  Brummelaar}, {Ridgway}, {Sturmann}, {Sturmann}, {Turner}, {Farrington}, \&
  {Goldfinger}}]{Ligi2012}
{Ligi}, R., {Mourard}, D., {Lagrange}, A.~M., {et~al.} 2012, \aap, 545, A5

\bibitem[{Ligi {et~al.}(2013)Ligi, Mourard, Nardetto, \& Clausse}]{Ligi2013}
Ligi, R., Mourard, D., Nardetto, N., \& Clausse, J.-M. 2013, Journal of
  Astronomical Instrumentation, 02, 1340003

\bibitem[{{Lindegren} {et~al.}(2018){Lindegren}, {Hern{\'a}ndez}, {Bombrun},
  {Klioner}, {Bastian}, {Ramos-Lerate}, {de Torres}, {Steidelm{\"u}ller},
  {Stephenson}, \& {Hobbs}}]{Lindegren2018}
{Lindegren}, L., {Hern{\'a}ndez}, J., {Bombrun}, A., {et~al.} 2018, \aap, 616,
  A2

\bibitem[{{Luck} \& {Heiter}(2005)}]{Luck2005}
{Luck}, R.~E. \& {Heiter}, U. 2005, \aj, 129, 1063

\bibitem[{{Luck} \& {Heiter}(2006)}]{Luck2006}
{Luck}, R.~E. \& {Heiter}, U. 2006, \aj, 131, 3069

\bibitem[{{Luri} {et~al.}(2018){Luri}, {Brown}, {Sarro}, {Arenou},
  {Bailer-Jones}, {Castro-Ginard}, {de Bruijne}, {Prusti}, {Babusiaux}, \&
  {Delgado}}]{Luri2018}
{Luri}, X., {Brown}, A.~G.~A., {Sarro}, L.~M., {et~al.} 2018, \aap, 616, A9

\bibitem[{{Maldonado} {et~al.}(2012){Maldonado}, {Eiroa}, {Villaver},
  {Montesinos}, \& {Mora}}]{Maldonado2012}
{Maldonado}, J., {Eiroa}, C., {Villaver}, E., {Montesinos}, B., \& {Mora}, A.
  2012, \aap, 541, A40

\bibitem[{{Maldonado} {et~al.}(2015){Maldonado}, {Eiroa}, {Villaver},
  {Montesinos}, \& {Mora}}]{Maldonado2015}
{Maldonado}, J., {Eiroa}, C., {Villaver}, E., {Montesinos}, B., \& {Mora}, A.
  2015, \aap, 579, A20

\bibitem[{{Mamajek} \& {Hillenbrand}(2008)}]{Mamajek2008}
{Mamajek}, E.~E. \& {Hillenbrand}, L.~A. 2008, \apj, 687, 1264

\bibitem[{{Mishenina} {et~al.}(2013){Mishenina}, {Pignatari}, {Korotin},
  {Soubiran}, {Charbonnel}, {Thielemann}, {Gorbaneva}, \&
  {Basak}}]{Mishenina2013}
{Mishenina}, T.~V., {Pignatari}, M., {Korotin}, S.~A., {et~al.} 2013, \aap,
  552, A128

\bibitem[{{Mishenina} {et~al.}(2012){Mishenina}, {Soubiran}, {Kovtyukh},
  {Katsova}, \& {Livshits}}]{Mishenina2012}
{Mishenina}, T.~V., {Soubiran}, C., {Kovtyukh}, V.~V., {Katsova}, M.~M., \&
  {Livshits}, M.~A. 2012, \aap, 547, A106

\bibitem[{{Mishenina} {et~al.}(2004){Mishenina}, {Soubiran}, {Kovtyukh}, \&
  {Korotin}}]{Mishenina2004}
{Mishenina}, T.~V., {Soubiran}, C., {Kovtyukh}, V.~V., \& {Korotin}, S.~A.
  2004, \aap, 418, 551

\bibitem[{{Morel} \& {Lebreton}(2008)}]{Morel2008}
{Morel}, P. \& {Lebreton}, Y. 2008, \apss, 316, 61

\bibitem[{{Motalebi} {et~al.}(2015){Motalebi}, {Udry}, {Gillon}, {Lovis},
  {S{\'e}gransan}, {Buchhave}, {Demory}, {Malavolta}, {Dressing}, {Sasselov},
  {Rice}, {Charbonneau}, {Collier Cameron}, {Latham}, {Molinari}, {Pepe},
  {Affer}, {Bonomo}, {Cosentino}, {Dumusque}, {Figueira}, {Fiorenzano},
  {Gettel}, {Harutyunyan}, {Haywood}, {Johnson}, {Lopez}, {Lopez-Morales},
  {Mayor}, {Micela}, {Mortier}, {Nascimbeni}, {Philips}, {Piotto}, {Pollacco},
  {Queloz}, {Sozzetti}, {Vanderburg}, \& {Watson}}]{Motalebi2015}
{Motalebi}, F., {Udry}, S., {Gillon}, M., {et~al.} 2015, \aap, 584, A72

\bibitem[{{Mourard} {et~al.}(2011){Mourard}, {B{\'e}rio}, {Perraut}, {Ligi},
  {Blazit}, {Clausse}, {Nardetto}, {Spang}, {Tallon-Bosc}, {Bonneau},
  {Chesneau}, {Delaa}, {Millour}, {Stee}, {Le Bouquin}, {ten Brummelaar},
  {Farrington}, {Goldfinger}, \& {Monnier}}]{Mourard2011}
{Mourard}, D., {B{\'e}rio}, P., {Perraut}, K., {et~al.} 2011, \aap, 531, A110

\bibitem[{{Mourard} {et~al.}(2012){Mourard}, {Challouf}, {Ligi}, {B{\'e}rio},
  {Clausse}, {Gerakis}, {Bourges}, {Nardetto}, {Perraut}, {Tallon-Bosc},
  {McAlister}, {ten Brummelaar}, {Ridgway}, {Sturmann}, {Sturmann}, {Turner},
  {Farrington}, \& {Goldfinger}}]{Mourard2012}
{Mourard}, D., {Challouf}, M., {Ligi}, R., {et~al.} 2012, in Society of
  Photo-Optical Instrumentation Engineers (SPIE) Conference Series, Vol. 8445

\bibitem[{{Mourard} {et~al.}(2009){Mourard}, {Clausse}, {Marcotto}, {Perraut},
  {Tallon-Bosc}, {B{\'e}rio}, {Blazit}, {Bonneau}, {Bosio}, {Bresson},
  {Chesneau}, {Delaa}, {H{\'e}nault}, {Hughes}, {Lagarde}, {Merlin}, {Roussel},
  {Spang}, {Stee}, {Tallon}, {Antonelli}, {Foy}, {Kervella}, {Petrov},
  {Thiebaut}, {Vakili}, {McAlister}, {ten Brummelaar}, {Sturmann}, {Sturmann},
  {Turner}, {Farrington}, \& {Goldfinger}}]{Mourard2009}
{Mourard}, D., {Clausse}, J.~M., {Marcotto}, A., {et~al.} 2009, \aap, 508, 1073

\bibitem[{{Mourard} {et~al.}(2015){Mourard}, {Monnier}, {Meilland}, {Gies},
  {Millour}, {Benisty}, {Che}, {Grundstrom}, {Ligi}, {Schaefer}, {Baron},
  {Kraus}, {Zhao}, {Pedretti}, {Berio}, {Clausse}, {Nardetto}, {Perraut},
  {Spang}, {Stee}, {Tallon-Bosc}, {McAlister}, {ten Brummelaar}, {Ridgway},
  {Sturmann}, {Sturmann}, {Turner}, \& {Farrington}}]{Mourard2015}
{Mourard}, D., {Monnier}, J.~D., {Meilland}, A., {et~al.} 2015, \aap, 577, A51

\bibitem[{{Nordlund} {et~al.}(2009){Nordlund}, {Stein}, \&
  {Asplund}}]{2009LRSP....6....2N}
{Nordlund}, {\r{A}}., {Stein}, R.~F., \& {Asplund}, M. 2009, Living Reviews in
  Solar Physics, 6, 2

\bibitem[{{Oja}(1993)}]{Oja1993}
{Oja}, T. 1993, Astronomy and Astrophysics Supplement Series, 100, 591

\bibitem[{{Pichierri} {et~al.}(2018){Pichierri}, {Morbidelli}, \&
  {Crida}}]{Pichierri+2018}
{Pichierri}, G., {Morbidelli}, A., \& {Crida}, A. 2018, Celestial Mechanics and
  Dynamical Astronomy, 130, 54

\bibitem[{{Prugniel} {et~al.}(2011){Prugniel}, {Vauglin}, \&
  {Koleva}}]{Prugniel2011}
{Prugniel}, P., {Vauglin}, I., \& {Koleva}, M. 2011, \aap, 531, A165

\bibitem[{{Ram{\'\i}rez} {et~al.}(2007){Ram{\'\i}rez}, {Allende Prieto}, \&
  {Lambert}}]{Ramirez2007}
{Ram{\'\i}rez}, I., {Allende Prieto}, C., \& {Lambert}, D.~L. 2007, \aap, 465,
  271

\bibitem[{{Ram{\'\i}rez} {et~al.}(2013){Ram{\'\i}rez}, {Allende Prieto}, \&
  {Lambert}}]{Ramirez2013}
{Ram{\'\i}rez}, I., {Allende Prieto}, C., \& {Lambert}, D.~L. 2013, \apj, 764,
  78

\bibitem[{{Ram{\'\i}rez} {et~al.}(2012){Ram{\'\i}rez}, {Fish}, {Lambert}, \&
  {Allende Prieto}}]{Ramirez2012}
{Ram{\'\i}rez}, I., {Fish}, J.~R., {Lambert}, D.~L., \& {Allende Prieto}, C.
  2012, \apj, 756, 46

\bibitem[{{Seager} \& {Mall{\'e}n-Ornelas}(2003)}]{Seager2003}
{Seager}, S. \& {Mall{\'e}n-Ornelas}, G. 2003, \apj, 585, 1038

\bibitem[{{Silva Aguirre} {et~al.}(2017){Silva Aguirre}, {Lund}, {Antia},
  {Ball}, {Basu}, {Christensen-Dalsgaard}, {Lebreton}, {Reese}, {Verma},
  {Casagrande}, {Justesen}, {Mosumgaard}, {Chaplin}, {Bedding}, {Davies},
  {Handberg}, {Houdek}, {Huber}, {Kjeldsen}, {Latham}, {White}, {Coelho},
  {Miglio}, \& {Rendle}}]{2017ApJ...835..173S}
{Silva Aguirre}, V., {Lund}, M.~N., {Antia}, H.~M., {et~al.} 2017, \apj, 835,
  173

\bibitem[{{Soubiran} {et~al.}(2008){Soubiran}, {Bienaym{\'e}}, {Mishenina}, \&
  {Kovtyukh}}]{Soubiran2008}
{Soubiran}, C., {Bienaym{\'e}}, O., {Mishenina}, T.~V., \& {Kovtyukh}, V.~V.
  2008, \aap, 480, 91

\bibitem[{Spaulding {et~al.}(2012)Spaulding, McWilliams, Jeanloz, Eggert,
  Celliers, Hicks, Collins, \& Smith}]{spaulding2012evidence}
Spaulding, D., McWilliams, R., Jeanloz, R., {et~al.} 2012, Physical Review
  Letters, 108, 065701

\bibitem[{{Stassun} \& {Torres}(2018)}]{Stassun2018}
{Stassun}, K.~G. \& {Torres}, G. 2018, \apj, 862, 61

\bibitem[{Stixrude \& Lithgow-Bertelloni(2011)}]{stixrude2011thermodynamics}
Stixrude, L. \& Lithgow-Bertelloni, C. 2011, Geophysical Journal International,
  184, 1180

\bibitem[{{Tallon-Bosc} {et~al.}(2008){Tallon-Bosc}, {Tallon}, {Thi{\'e}baut},
  {B{\'e}chet}, {Mella}, {Lafrasse}, {Chesneau}, {Domiciano de Souza},
  {Duvert}, {Mourard}, {Petrov}, \& {Vannier}}]{Tallon2008}
{Tallon-Bosc}, I., {Tallon}, M., {Thi{\'e}baut}, E., {et~al.} 2008, in Society
  of Photo-Optical Instrumentation Engineers (SPIE) Conference Series, Vol.
  7013

\bibitem[{{ten Brummelaar} {et~al.}(2005){ten Brummelaar}, {McAlister},
  {Ridgway}, {Bagnuolo}, {Turner}, {Sturmann}, {Sturmann}, {Berger}, {Ogden},
  {Cadman}, {Hartkopf}, {Hopper}, \& {Shure}}]{tenBrummelaar2005}
{ten Brummelaar}, T.~A., {McAlister}, H.~A., {Ridgway}, S.~T., {et~al.} 2005,
  \apj, 628, 453

\bibitem[{{Valenti} \& {Fischer}(2005)}]{Valenti2005}
{Valenti}, J.~A. \& {Fischer}, D.~A. 2005, \apjs, 159, 141

\bibitem[{{Vogt} {et~al.}(2015){Vogt}, {Burt}, {Meschiari}, {Butler}, {Henry},
  {Wang}, {Holden}, {Gapp}, {Hanson}, {Arriagada}, {Keiser}, {Teske}, \&
  {Laughlin}}]{Vogt2015}
{Vogt}, S.~S., {Burt}, J., {Meschiari}, S., {et~al.} 2015, \apj, 814, 12

\bibitem[{{Wenger} {et~al.}(2000){Wenger}, {Ochsenbein}, {Egret}, {Dubois},
  {Bonnarel}, {Borde}, {Genova}, {Jasniewicz}, {Lalo{\"e}}, {Lesteven}, \&
  {Monier}}]{CDS}
{Wenger}, M., {Ochsenbein}, F., {Egret}, D., {et~al.} 2000, \aaps, 143, 9

\bibitem[{Wolf \& Bower(2018)}]{wolf2018equation}
Wolf, A.~S. \& Bower, D.~J. 2018, Physics of the Earth and Planetary Interiors,
  278, 59

\end{thebibliography}

\begin{appendix}
\section{Selected \logg and metallicity from literature.}

\begin{table}[h]
    \centering
       \caption{Parameters used to derive the \logg and \FeH of Table~\ref{tab:StellarParam}.}
    \begin{tabular}{l|c|c|l}
    \hline \hline
\Teff [K] & \logg [dex] & \FeH    &  Reference     \\
\hline
  5100 & 4.40  &  0.10   &  \cite{Heiter2003} \\ 
  5100 & 4.65  &  0.04   &  \cite{Luck2006} \\ 
  5100 & 4.65  &  0.04   &  \cite{Luck2005} \\ 
  4798 & 4.55  &   -     &  \cite{Gray2003} \\ 
  4732 & 4.37  &  0.09   & \cite{Boeche2016} \\
  4858 & 4.67  &  -     &  \cite{Maldonado2015} \\
  5044 & 4.58  & 0.04   &  \cite{DaSilva2015} \\ 
  4900 & 4.20  & 0.05   &  \cite{Mishenina2013} \\
  4833 & 4.59  & 0.00   &  \cite{Ramirez2013} \\ 
  4889 & 4.60  & 0.10   &  \cite{Mishenina2012} \\
  4833 & 4.59  & 0.00   &  \cite{Ramirez2012} \\ 
   -   &  -    & 0.10   &  \cite{Maldonado2012} \\ 
  4851 & 4.37  & 0.07   &  \cite{Lee2011} \\ 
  4715 & 4.57  & 0.06   &  \cite{Prugniel2011}\\ 
  4710 & 4.50  & 0.20   &  \cite{Frasca2009} \\ 
  4913 & 4.51  & 0.08   &  \cite{Soubiran2008}\\ 
  4825 & 4.62  & 0.05   &  \cite{Ramirez2007} \\ 
  4835 & 4.56  & 0.12   &  \cite{Valenti2005} \\ 
  4900 & 4.20  & 0.05   &  \cite{Mishenina2004} \\ 
  4743 & 4.63  & 0.12   &  \cite{Allende2004} \\ 
  \hline
    \end{tabular}
    \tablefoot{We took into account the values given in the CDS database, removing those which were redundant and obtained before 2000.}
    \label{tab:ParamLit}
\end{table}

\end{appendix}

\end{document}